\documentclass[prd,aps,showpacs,nofootinbib,preprint,eqsecnum]{revtex4}
\usepackage{graphicx,color,amsmath,amsxtra}
\usepackage{epsf}
\usepackage{amssymb}
\usepackage{enumerate}
\usepackage{hhline}
\usepackage{array}
\usepackage{tabularx}

\newcommand{\bear}{\begin{array}}  \newcommand{\eear}{\end{array}}
\newcommand{\bea}{\begin{eqnarray}}  \newcommand{\eea}{\end{eqnarray}}
\newcommand{\beq}{\begin{equation}}  \newcommand{\eeq}{\end{equation}}
\newcommand{\bef}{\begin{figure}}  \newcommand{\eef}{\end{figure}}
\newcommand{\bec}{\begin{center}}  \newcommand{\eec}{\end{center}}

  \newcommand{\abs}[1]{\vert{#1}\vert}

\newcommand{\Eqn}[1]{&\hspace{-0.2em}#1\hspace{-0.2em}&}
\def\be{\begin{equation}}
\def\ee{\end{equation}}
\def\bea{\begin{eqnarray}}
\def\eea{\end{eqnarray}}
\def\beq{\begin{eqnarray}}
\def\eeq{\end{eqnarray}}

 \def\be{\begin{equation}}
\def\ee{\end{equation}}
\def\bea{\begin{eqnarray}}
\def\eea{\end{eqnarray}}
\def\beq{\begin{eqnarray}}
\def\eeq{\end{eqnarray}}

\begin{document}

\title{Reconstruction of $f(T)$ gravity: Rip cosmology, finite-time future singularities and thermodynamics}

\author{Kazuharu Bamba$^{1, }$\footnote{
E-mail address: bamba@kmi.nagoya-u.ac.jp}, 
Ratbay Myrzakulov$^{2, }$\footnote{
E-mail addresses: rmyrzakulov@csufresno.edu; rmyrzakulov@gmail.com}, 
Shin'ichi Nojiri$^{1, 3, }$\footnote{E-mail address:
nojiri@phys.nagoya-u.ac.jp} 
and 
Sergei D. Odintsov$^{3, 4, 5, 6,}$\footnote{
E-mail address: odintsov@ieec.uab.es}}
\affiliation{
$^1$Kobayashi-Maskawa Institute for the Origin of Particles and the
Universe,
Nagoya University, Nagoya 464-8602, Japan\\
$^2$Eurasian International Center for Theoretical Physics, Eurasian National University, Astana 010008, Kazakhstan\\ 
$^3$Department of Physics, Nagoya University, Nagoya 464-8602, Japan\\
$^4$Instituci\`{o} Catalana de Recerca i Estudis Avan\c{c}ats (ICREA), 
Barcelona, Spain\\ 
$^5$Institut de Ciencies de l'Espai (CSIC-IEEC),
Campus UAB, Facultat de Ciencies, Torre C5-Par-2a pl, E-08193 Bellaterra
(Barcelona), Spain\\ 
$^6$Tomsk State Pedagogical University, Tomsk, Russia
}

%\date{\today}

%%%%%%%%%%%%%%%%%%%%%
%  Abstract
%%%%%%%%%%%%%%%%%%%%%
\begin{abstract}

We demonstrate that there appear finite-time future singularities in 
$f(T)$ gravity with $T$ being the torsion scalar. 
We reconstruct a model of $f(T)$ gravity 
with realizing the finite-time future singularities. 
In addition, it is explicitly shown that 
a power-low type correction term $T^\beta$ ($\beta>1$) such as 
a $T^2$ term can remove the finite-time future singularities in $f(T)$ 
gravity. 
Moreover, we study $f(T)$ models with realizing 
inflation in the early universe, 
the $\Lambda$CDM model, Little Rip cosmology and 
Pseudo-Rip cosmology. 
It is demonstrated that the disintegration of bound structures for 
Little Rip and Pseudo-Rip cosmologies occurs in the same way as in 
gravity with corresponding dark energy fluid. 
We also discuss that the time-dependent matter instability in the star 
collapse can occur in $f(T)$ gravity. 
Furthermore, we explore thermodynamics in $f(T)$ gravity and 
illustrate that the second law of thermodynamics 
can be satisfied around the finite-time future singularities 
for the universe with the temperature inside the horizon being the same 
as that of the apparent horizon. 

\end{abstract}

%%%%%%%%%%%%%%%%%%%%%

%----------------------------
%\pacs{
%Keywords:
%}
%\preprint{}
%\hspace{13.0cm}
%----------------------------

\maketitle
%==============================================================================

%%%%%%%%%%%%%%%%%%%%%%%%%%%
%%%  Sec. I
%%%%%%%%%%%%%%%%%%%%%%%%%%%
\section{Introduction}

A number of cosmological observations, e.g., 
Type Ia Supernovae~\cite{SN1}, 
cosmic microwave background (CMB) radiation~\cite{WMAP, 
%Komatsu:2008hk, Komatsu:2010fb
Komatsu-WMAP}, 
large scale structure (LSS)~\cite{LSS}, 
baryon acoustic oscillations (BAO)~\cite{Eisenstein:2005su}, 
and weak lensing~\cite{Jain:2003tba}, 
have been implied the current accelerating expansion of the universe. 
%%%
Approaches to account for the late time cosmic acceleration 
are classified into two representative categories. 
The first is the introduction of 
unknown matters, i.e., so-called ``dark energy'', 
in the framework of general relativity (for recent reviews, see~\cite{
%Copeland:2006wr, Tsujikawa:2010sc, 
Book-Amendola-Tsujikawa, Li:2011sd}). 
The second is the modification of gravity such as  
$f(R)$ gravity (for recent reviews, see~\cite{Review-Nojiri-Odintsov, 
Book-Capozziello-Faraoni, Capozziello:2011et, 
DeFelice:2010aj, Clifton:2011jh}). 

As a gravitational theory beyond general relativity, 
one could consider ``teleparallelism" with 
the Weitzenb\"{o}ck connection, which has torsion $T$ and 
not the curvature $R$ defined by the Levi-Civita 
connection~\cite{Teleparallelism}. 
%It is known that Einstein also taken this procedure~\cite{Einstein}. 
In modern cosmology, 
in order to explain both inflation~\cite{Inflation-F-F} and the late time 
accelerated expansion of the universe, 
the teleparallel Lagrangian density represented by the torsion scalar $T$ 
has been extended to a function of $T$ as 
$f(T)$~\cite{Bengochea:2008gz, Linder:2010py}. 
This idea is equivalent to the concept of $f(R)$ gravity. 
%%%%%
In the recent literature, 
to check whether $f(T)$ gravity can be an alternative gravitational theory 
to general relativity, its various properties have been diversely 
explored~\cite{f(T)-Refs, BGL-Comment, Zheng:2010am, Local-Lorentz-invariance, 
%Ferraro:2011us, 
Thermodynamics-f(T), Bamba:2011pz}, e.g., the local Lorentz 
invariance~\cite{Local-Lorentz-invariance}, non-trivial conformal frames and 
thermodynamics~\cite{Thermodynamics-f(T), Bamba:2011pz}. 

Moreover, 
it is known that if (phantom/quintessence) dark energy dominates the universe, 
in general there can appear finite-time future singularities, 
which have been classified into four types~\cite{Nojiri:2005sx}. 
The finite-time future singularities 
in $f(R)$ gravity~\cite{FS-F(R)-gravity} have first been observed and 
those in various modified gravity~\cite{Future-singularity-MG, Bamba:2011ky} 
have also been investigated. 
Therefore, it is important to examine models of $f(T)$ gravity in which 
finite-time future singularities can exist. 
%%%%%

%%%
In this paper, we concentrate on the two important theoretical features of 
$f(T)$ gravity: the finite-time future singularities and 
thermodynamics in $f(T)$ gravity. 
First, we explicitly reconstruct $f(T)$ gravity 
in which the finite-time future singularities appear by following the 
procedure proposed in Refs.~\cite{Reconstruction-Method, 
%Nojiri:2006gh, Nojiri:2006be, Nojiri:2008fk, 
Nojiri:2009kx, Nojiri:2011kd}. 
We also study a correction term to the models of $f(T)$ gravity, 
%which 
so that such a term 
can remove the finite-time future singularities 
in analogy with $f(R)$ gravity. 
%%%%%
Moreover, we explore the reconstruction of $f(T)$ models which 
realize the examples of inflation in the early universe, 
the $\Lambda$CDM model, 
Little Rip cosmology~\cite{Frampton:2011sp, Brevik:2011mm, 
%FLNOS-ANOY, 
Frampton:2011rh, Astashenok:2012tv, 
Nojiri:2011kd, GL-IT, Ito:2011ae, BBEO-XZL-MKO} and 
Pseudo-Rip cosmology~\cite{Frampton:2011aa}. 
The Little Rip scenario is a kind of a mild phantom scenario and 
considered in order to avoid a Big Rip singularity. 
%%%
On the other hand, the Pseudo-Rip model is an intermediate case between 
the cosmological constant and the Little Rip cosmology. In this model, 
the Hubble parameter asymptotically becomes constant as time goes to 
infinity, although for the Big Rip singularity, the Hubble parameter diverges 
at finite time, and in the Little Rip cosmology the Hubble parameter becomes 
infinity asymptotically as time goes to infinity. 
%%%
%%%%%
Furthermore, 
we examine whether the time-dependent matter instability in the star 
collapse~\cite{Arbuzova:2010iu} occurs in $f(T)$ gravity in analogy with $f(R)$ gravity. This instability has recently been found in the framework of 
$f(R)$ gravity, in addition to the well-known matter 
instability~\cite{Dolgov:2003px}. 
Next, we explore thermodynamics in $f(T)$ gravity, especially, 
near to the finite-time future singularities. 
In particular, we demonstrate that the second law of thermodynamics 
can be satisfied around the finite-time future singularities 
if the temperature of the universe inside the 
horizon is the same as that of the apparent horizon. 
%%%
%%%%% Units %%%%%
We use units of $k_\mathrm{B} = c = \hbar = 1$ and denote the
gravitational constant $8 \pi G$ by 
${\kappa}^2 \equiv 8\pi/{M_{\mathrm{Pl}}}^2$ 
with the Planck mass of $M_{\mathrm{Pl}} = G^{-1/2} = 1.2 \times 10^{19}$GeV.
%%%%%%%%%%%%%%%%%

%%%%% Structure %%%%%
The paper is organized as follows. 
In Sec.\ II, we explain the fundamental formulations and basic equations 
in $f(T)$ gravity. 
In Sec.\ III, we investigate the finite-time future singularities, 
using analogy with $f(R)$ gravity. 
In Sec.\ IV, we reconstruct $f(T)$ gravity models with 
realizing the finite-time future singularities. 
We also study a correction term which can remove 
the finite-time future singularities. 
In addition, 
we reconstruct $f(T)$ models with realizing 
inflation in the early universe, 
the $\Lambda$CDM model, Little Rip cosmology and Pseudo-Rip cosmology. 
The calculation of an inertial force which may lead to the dissolution of 
bound structures is done in Little Rip and Pseudo-Rip cosmologies 
for the Earth-Sun (ES) system. 
Furthermore, 
we discuss that the time-dependent matter instability in the star collapse can 
occur in $f(T)$ gravity in analogy with $f(R)$ gravity. 
In Sec.\ V, we explore thermodynamics in $f(T)$ gravity. 
We demonstrate that the second law of thermodynamics 
can be satisfied around the finite-time future singularities. 
Finally, conclusions are given in Sec.\ VI. 
%%%%%%%%%%%%%%%%%%%%%

%%%%%%%%%%%%%%%%%%%%%%%%%%%
%%%  Sec. II
%%%%%%%%%%%%%%%%%%%%%%%%%%%
\section{$f(T)$ gravity}

%%%%%%%%%%%%%%%%%%%%%%%%%%%
%%%  Sec. II A
%%%%%%%%%%%%%%%%%%%%%%%%%%%
\subsection{Fundamental formulations}

Orthonormal tetrad components $e_A (x^{\mu})$ are used in the teleparallelism. 
An index $A$ runs over $0, 1, 2, 3$ for the 
tangent space at each point $x^{\mu}$ of the manifold. 
Their relation to the metric $g^{\mu\nu}$ is described as 
%
%\begin{equation}
$
g_{\mu\nu}=\eta_{A B} e^A_\mu e^B_\nu 
$. 
%\label{eq:2.1}
%\end{equation}
%
Here, $\mu$ and $\nu$ are coordinate indices on the manifold, 
which also run over $0, 1, 2, 3$, 
and $e_A^\mu$ forms the tangent vector of the manifold. 
The torsion $T^\rho_{\verb| |\mu\nu}$ and contorsion 
$K^{\mu\nu}_{\verb|  |\rho}$ tensors are defined as 
\begin{eqnarray}
T^\rho_{\verb| |\mu\nu} \Eqn{\equiv} e^\rho_A 
\left( \partial_\mu e^A_\nu - \partial_\nu e^A_\mu \right)\,, 
\label{eq:2.2} \\ 
K^{\mu\nu}_{\verb|  |\rho} 
\Eqn{\equiv} 
-\frac{1}{2} 
\left(T^{\mu\nu}_{\verb|  |\rho} - T^{\nu \mu}_{\verb|  |\rho} - 
T_\rho^{\verb| |\mu\nu}\right)\,. 
\label{eq:2.3}
\end{eqnarray}
The teleparallel Lagrangian density is expressed by using the torsion scalar 
$T$, although in general relativity the Lagrangian 
density is described by the Ricci scalar $R$. The torsion scalar 
$T$ is given by
\begin{eqnarray}
T \Eqn{\equiv} S_\rho^{\verb| |\mu\nu} T^\rho_{\verb| |\mu\nu}\,,
\label{eq:2.4} \\ 
%\end{equation}
%
%with
%
%\begin{equation}
S_\rho^{\verb| |\mu\nu} \Eqn{\equiv} \frac{1}{2}
\left(K^{\mu\nu}_{\verb|  |\rho}+\delta^\mu_\rho \ 
T^{\alpha \nu}_{\verb|  |\alpha}-\delta^\nu_\rho \ 
T^{\alpha \mu}_{\verb|  |\alpha}\right)\,. 
\label{eq:2.5}
\end{eqnarray}
The modified teleparallel action describing $f(T)$ 
gravity~\cite{Linder:2010py} is as follows: 
\begin{equation}
I= 
\int d^4x \abs{e} \left[ \frac{f(T)}{2{\kappa}^2} 
+{\mathcal{L}}_{\mathrm{M}} \right]\,,
\label{eq:2.6}
\end{equation}
where $\abs{e}= \det \left(e^A_\mu \right)=\sqrt{-g}$ and 
${\mathcal{L}}_{\mathrm{M}}$ is the Lagrangian of matter. 
%%%%%
The variation of the action in Eq.~(\ref{eq:2.6}) with respect to 
the vierbein vector field $e_A^\mu$ presents~\cite{Bengochea:2008gz} 
\begin{equation}
\frac{1}{e} \partial_\mu \left( eS_A^{\verb| |\mu\nu} \right) f^{\prime} 
-e_A^\lambda T^\rho_{\verb| |\mu \lambda} S_\rho^{\verb| |\nu\mu} 
f^{\prime} +S_A^{\verb| |\mu\nu} \partial_\mu \left(T\right) f^{\prime\prime} 
+\frac{1}{4} e_A^\nu f = \frac{{\kappa}^2}{2} e_A^\rho 
{T^{(\mathrm{M})}}_\rho^{\verb| |\nu}\,, 
\label{eq:2.7}
\end{equation}
where ${T^{(\mathrm{M})}}_\rho^{\verb| |\nu}$ 
is the energy-momentum tensor of all 
perfect fluids of ordinary matter, i.e., 
radiation and non-relativistic matter. 
%%%%%

%%%%%%%%%%%%%%%%%%%%%%%%%%%
%%%  Sec. II B
%%%%%%%%%%%%%%%%%%%%%%%%%%%
\subsection{Basic equations}

%%%%%
We take the four-dimensional flat 
Friedmann-Lema\^{i}tre-Robertson-Walker (FLRW) 
space-time with the metric 
\begin{equation}
d s^2 = h_{\alpha \beta} d x^{\alpha} d x^{\beta}
+\tilde{r}^2 d \Omega^2\,. 
\label{eq:2.8}
\end{equation}
Here, $\tilde{r}=a(t)r$, $x^0=t$ and $x^1=r$ with the two-dimensional 
metric $h_{\alpha \beta}={\rm diag}(1, -a^2(t))$, 
$a(t)$ is the scale factor, and 
$d \Omega^2$ is the metric of two-dimensional sphere with unit radius. 
%%%%%
In this background, we have 
$g_{\mu \nu}= \mathrm{diag} (1, -a^2, -a^2, -a^2)$ and 
the tetrad components $e^A_\mu = (1,a,a,a)$. 
By using these relations, 
we find the exact value of torsion scalar $T=-6H^2$ with 
$H = \dot{a}/a$ being the Hubble parameter, 
where the dot denotes the time derivative, $\partial/\partial t$. 

In the flat FLRW background, 
the gravitational field equations can be written in the equivalent forms of 
those in general relativity: 
\begin{eqnarray}
H^2 \Eqn{=} \frac{{\kappa}^2}{3} \left(\rho_{\mathrm{M}}+\rho_{\mathrm{DE}} 
\right)\,, 
%\label{2-4}
\label{eq:4.1} \\ 
%
%\frac{d H^2}{d \ln a} 
%
\dot{H}
\Eqn{=} -\frac{{\kappa}^2}{2} \left(\rho_{\mathrm{M}} + P_{\mathrm{M}} + 
\rho_{\mathrm{DE}} + P_{\mathrm{DE}} \right)\,,
%\label{2-5}
\label{eq:4.2} 
\end{eqnarray}
where $F\equiv df/dT$, $F^{\prime}=dF/dT$, and 
$\rho_{\mathrm{M}}$ and $P_{\mathrm{M}}$ are 
the energy density and pressure of all perfect fluids of generic matter, 
respectively. 
%%%%%
The perfect fluid satisfies the continuity equation 
% 
%\begin{equation}
$
\dot{\rho}_{\mathrm{M}}+3H\left( \rho_{\mathrm{M}} + P_{\mathrm{M}} \right)
=0 
$.
%\label{eq:2.11}
%\end{equation}
%
%%%%%
Moreover, 
the energy density and pressure of dark components 
can be represented by 
\begin{eqnarray}
\rho_{\mathrm{DE}} 
\Eqn{=} 
\frac{1}{2{\kappa}^2} J_1\,,
%\left( -T -f +2TF \right)\,, 
\label{eq:4.3} \\ 
P_{\mathrm{DE}} 
\Eqn{=} 
-\frac{1}{2{\kappa}^2} 
%\left[ 
\left( 
4J_2 
%4J_2 \dot{H} 
%4\left(1 -F -2TF^{\prime} \right) \dot{H} 
+ J_1 
%\left( -T -f +2TF \right)
\right)\,,
%\right]\,, 
\label{eq:4.4} 
\end{eqnarray}
with 
\begin{eqnarray}
J_1 \Eqn{\equiv} -T -f +2TF\,, 
\label{eq:IIB-Add-01} \\ 
J_2 \Eqn{\equiv} \left( 1 -F -2TF^{\prime} \right) \dot{H}\,. 
\label{eq:IIB-Add-02}
\end{eqnarray}
Here, $\rho_{\mathrm{DE}}$ in Eq.~(\ref{eq:4.3}) and $P_{\mathrm{DE}}$ 
in Eq.~(\ref{eq:4.4}) satisfy the standard continuity equation 
\begin{equation} 
\dot{\rho}_{\mathrm{DE}}+3H \left( 
\rho_{\mathrm{DE}} + P_{\mathrm{DE}}
\right)  
= 0\,. 
\label{eq:4.5}
\end{equation} 
%

%%%%%%%%%%%%%%%%%%%
%%%  Sec. III
%%%%%%%%%%%%%%%%%%%
\section{Finite-time future singularities in $f(T)$ gravity}

%%%%%%%%%%%%%%%%%%%
%%%  Sec. III A
%%%%%%%%%%%%%%%%%%% 
\subsection{Classification of the four types}

In the FLRW background~(\ref{eq:2.2}), 
the effective equation of state (EoS) 
for the universe is given by~\cite{Review-Nojiri-Odintsov} 
\begin{eqnarray} 
w_{\mathrm{eff}} \Eqn{\equiv} \frac{P_{\mathrm{eff}}}{\rho_{\mathrm{eff}}} = 
-1 - \frac{2\dot{H}}{3H^2}\,, 
\label{eq:2.17} \\
\rho_{\mathrm{eff}} \Eqn{\equiv} \frac{3H^2}{\kappa^2}\,, 
\label{eq:III-add-01} \\
P_{\mathrm{eff}} \Eqn{\equiv} -\frac{2\dot{H}+3H^2}{\kappa^2}\,. 
\label{eq:III-add-02} 
\end{eqnarray}
Here, 
$\rho_{\mathrm{eff}}$ and $P_{\mathrm{eff}}$ correspond to 
the total energy density and pressure of the universe, respectively. 
%%%
When the energy density of dark energy becomes completely dominant over 
that of matter, one can consider 
%$w_{\mathrm{eff}} \approx w_{\mathrm{DE}}$. 
$w_{\mathrm{DE}} \approx w_{\mathrm{eff}}$.
%%%%%
For $\dot{H} < 0\ (>0)$, $w_\mathrm{eff} >-1\ (<-1)$, representing the non-phantom, i.e., quintessence (phantom) phase, 
whereas 
$w_\mathrm{eff} =-1$ for $\dot{H} = 0$, corresponding to the 
cosmological constant. 
%%%%%

%%%%%
%It is  known that 
In Ref.~\cite{Nojiri:2005sx}, 
the finite-time future singularities has been classified into the following 
four types. 
%
%\begin{itemize}
%\item 
(i) 
Type I (``Big Rip''~\cite{Big-Rip}):\ 
In the limit $t\to t_{\mathrm{s}}$, 
$a \to \infty$,
$\rho_{\mathrm{eff}} \to \infty$ and
$\abs{ P_{\mathrm{eff}} } \to \infty$. 
The case in which 
$\rho_\mathrm{{eff}}$ and $P_{\mathrm{eff}}$ becomes finite values 
at $t = t_{\mathrm{s}}$~\cite{Shtanov:2002ek} 
is also included. 
%
%\item 
(ii) 
Type II (``sudden''~\cite{Barrow:2004xh, sudden}):\ 
In the limit $t\to t_{\mathrm{s}}$, 
$a \to a_{\mathrm{s}}$, 
$\rho_{\mathrm{eff}} \to \rho_{\mathrm{s}}$ and 
$\abs{ P_{\mathrm{eff}} } \to \infty$. 
%
%\item 
(iii)
Type III:\ 
In the limit $t\to t_{\mathrm{s}}$, 
$a \to a_{\mathrm{s}}$, 
$\rho_{\mathrm{eff}} \to \infty$ and
$\abs{ P_{\mathrm{eff}} } \to \infty$. 
%
%\item 
(iv) 
Type IV:\ 
In the limit $t\to t_{\mathrm{s}}$, 
$a \to a_{\mathrm{s}}$, 
$\rho_{\mathrm{eff}} \to 0$, 
$\abs{ P_{\mathrm{eff}} } \to 0$, 
and higher derivatives of $H$ diverge. 
The case in which $\rho_{\mathrm{eff}}$ and/or $\abs{ P_{\mathrm{eff}} }$ 
asymptotically approach finite values is also included. 
%\end{itemize}
%
%%%%%
Here, $t_{\mathrm{s}}$, $a_{\mathrm{s}} (\neq 0)$ and $\rho_{\mathrm{s}}$ 
are constants. 

%%%%% 
It is important to mention that the 
Type I, i.e., ``Big Rip'' singularity, has recently been extended by 
Little Rip~\cite{Frampton:2011sp, Brevik:2011mm, 
Frampton:2011rh, Astashenok:2012tv, 
Nojiri:2011kd, GL-IT, Ito:2011ae, BBEO-XZL-MKO} and Pseudo Rip~\cite{Frampton:2011aa} scenarios. 
Furthermore, 
in addition to the Type V (``$w$'') singularity, (v) Type V 
(``$w$''~\cite{Kiefer:2010zzb, Dabrowski:2009zzb, Dabrowski:2009pc}) 
singularity and earlier by parallel-propagated (p.p.) curvature 
singularities~\cite{L-Fernandez-Jambrina-PRD} have now been proposed. 
For the Type V (``$w$'') singularity, 
in the limit $t\to t_{\mathrm{s}}$, 
$a \to a_{\mathrm{s}}$, 
$\rho_{\mathrm{eff}} \to 0$, 
$\abs{ P_{\mathrm{eff}} } \to 0$, 
and the EoS for the universe diverges. 
It should be cautioned that the Type V (``$w$'') singularity is similar to 
the Type IV singularity, namely, the Type V singularity does not lead 
to the divergence of physical quantities such as the scale factor, the energy density and pressure in the future, 
but in the Type V higher derivatives of $H$ do not diverge. 
This is the difference between the Type IV and Type V singularities. 
%%%%%

In what follows, since we examine the behavior of the universe around 
the finite-time future singularities, in which 
dark energy dominates over matter as 
$\rho_{\mathrm{DE}} \gg \rho_{\mathrm{M}}$ and 
$P_{\mathrm{DE}} \gg P_{\mathrm{M}}$ in Eqs.~(\ref{eq:4.1}) and 
(\ref{eq:4.2}), 
we consider $\rho_{\mathrm{eff}} \approx \rho_{\mathrm{DE}}$ in 
Eq.~(\ref{eq:4.3}), 
$P_{\mathrm{eff}} \approx P_{\mathrm{DE}}$ in 
Eq.~(\ref{eq:4.4}), and 
$w_{\mathrm{eff}} \approx w_{\mathrm{DE}} = 
P_{\mathrm{DE}}/\rho_{\mathrm{DE}}$.
%%%%%

%%%%%%%%%%%%%%%%%%%
%%%  Sec. III B
%%%%%%%%%%%%%%%%%%%
\subsection{Finite-time future singularities}

We consider the case in which the Hubble parameter 
$H$ is represented by 
[the third reference in Ref.~\cite{Reconstruction-Method}] 
\begin{eqnarray} 
H \Eqn{\sim} \frac{h_{\mathrm{s}}}{ \left( t_{\mathrm{s}} - t 
\right)^{q}}\,, 
\quad 
\mathrm{for} \,\,\, q > 0\,,
\label{eq:2.13} \\ 
H \Eqn{\sim} H_{\mathrm{s}} + \frac{h_{\mathrm{s}}}{ 
\left( t_{\mathrm{s}} - t \right)^{q}}\,, 
\quad 
\mathrm{for} \,\,\, q<-1\,, \,\,\, -1< q < 0\,. 
\label{eq:IIIB-add-001}
\end{eqnarray}
Here, 
$h_{\mathrm{s}} (> 0)$ and $H_{\mathrm{s}} (> 0)$ are positive constants, 
$q (\neq 0, \, -1)$ is a non-zero constant. 
Moreover, $t_{\mathrm{s}}$ is the time when the finite-time future singularity 
appears and only the period $0< t < t_{\mathrm{s}}$ is considered 
due to the fact that $H$ should be real number. 
%%%
%%%%%
When $t\to t_{\mathrm{s}}$, 
for $q > 0$, 
%$q>1$, 
both 
$H \sim h_{\mathrm{s}} \left( t_{\mathrm{s}} - t 
\right)^{-q}$ 
and 
$\dot{H} \sim q h_{\mathrm{s}} \left( t_{\mathrm{s}} - t 
\right)^{-\left(q+1\right)}$ become infinity. 
For $-1 < q < 0$, $H$ is finite, 
but $\dot{H}$ becomes infinity. 
For $q < -1$, but $q$ is not any integer, both $H$ and $\dot{H}$ are finite, 
but the higher derivatives of $H$ can become infinity. 
%%%%%
It follows from Eq.~(\ref{eq:2.13}) that 
\begin{eqnarray} 
a \Eqn{\sim} a_{\mathrm{s}} \exp \left[ \frac{h_{\mathrm{s}}}{q-1} 
\left( t_{\mathrm{s}} - t 
\right)^{-\left(q-1\right)}
\right]
\quad 
\mathrm{for}\,\,\, 
%q \neq 1\,,
%-1<q<0\,, \,\,\, 
0<q<1\,, \,\,\, 1<q\,, 
\label{eq:2.14} \\
a \Eqn{\sim} a_{\mathrm{s}} \frac{h_{\mathrm{s}}}{\left( t_{\mathrm{s}} - t 
\right)^{h_{\mathrm{s}}}}
\quad 
\mathrm{for}\,\,\, q = 1\,,
\label{eq:IIIB-add-01}
\end{eqnarray}
where $a_{\mathrm{s}}$ is a constant. 

It can be seen from Eq.~(\ref{eq:2.14}) that 
when $t\to t_{\mathrm{s}}$, 
for $q \geq 1$, 
%for $q>1$, 
$a \to \infty$, 
whereas 
for $q < 0$ and $0 < q < 1$, 
$a \to a_{\mathrm{s}}$. 
Moreover, 
it follows from Eqs.~(\ref{eq:III-add-01}) and (\ref{eq:2.13}) that 
for $q > 0$, 
$H \to \infty$ and therefore 
$\rho_{\mathrm{eff}} = 3 H^2/\kappa^2 \to \infty$, 
whereas 
for $q < 0$, $H$ asymptotically becomes finite 
and also $\rho_{\mathrm{eff}}$ asymptotically approaches a finite 
constant value $\rho_{\mathrm{s}}$. 
On the other hand, from 
$\dot{H} \sim q h_{\mathrm{s}} \left( t_{\mathrm{s}} - t 
\right)^{-\left(q+1\right)}$ and Eq.~(\ref{eq:III-add-02}) 
we find that 
for $q>-1$, $\dot{H} \to \infty$ and hence 
$P_{\mathrm{eff}} = -\left(2\dot H + 3H^2\right)/\kappa^2 \to \infty$. 
For $q < -1$, but $q$ is not any integer, 
$a$, $\rho_{\mathrm{eff}}$ and $P_{\mathrm{eff}}$ are finite
because both $H$ and $\dot{H}$ are finite, whereas the higher derivatives of 
$H$ diverges. 
%%%
As a result, the properties of the finite-time future singularities 
described by the expressions of $H$ 
in Eqs.~(\ref{eq:2.13}) and (\ref{eq:IIIB-add-001}) are summarized as follows: 
For $q \geq 1$, 
the Type I (``Big Rip'') singularity, 
for $0<q<1$, the Type III singularity, 
and for $-1<q<0$, the Type II (``sudden'') singularity. 
In addition, 
for $q < -1$, but $q$ is not any integer, 
the Type IV singularity appears. 
We present the conditions for the finite-time future singularities 
to exist on $q$ in the expressions of $H$ in Eqs.~(\ref{eq:2.13}) and 
(\ref{eq:IIIB-add-001}), $\rho_{\mathrm{DE}}$ in Eq.~(\ref{eq:4.3}) 
and $P_{\mathrm{DE}}$ in Eq.~(\ref{eq:4.4}), 
and the behaviors of $H$ and $\dot{H}$ 
in the limit of $t \to t_{\mathrm{s}}$ 
in Table \ref{tb:table1}.

%%%%%%%%% Table I %%%%%%%%%%%
\begin{table*}[tbp]
\caption{Conditions for the finite-time future singularities 
to exist on $q$ in the expressions of $H$ in Eqs.~(\ref{eq:2.13}) and 
(\ref{eq:IIIB-add-001}), $\rho_{\mathrm{DE}}$ in Eq.~(\ref{eq:4.3}) 
and $P_{\mathrm{DE}}$ in Eq.~(\ref{eq:4.4}), 
and the behaviors of $H$ and $\dot{H}$ 
in the limit of $t \to t_{\mathrm{s}}$. 
}
\begin{center}
\begin{tabular}
{lllll}
\hline
\hline
$q (\neq 0, \, -1)$
%& Behavior of $H$ in the limit of $t \to t_{\mathrm{s}}$ 
%& Behavior of $\dot{H}$ in the limit of $t \to t_{\mathrm{s}}$ 
& $H$ ($t \to t_{\mathrm{s}}$) \quad 
& $\dot{H}$ ($t \to t_{\mathrm{s}}$)
& $\rho_{\mathrm{DE}}$
& $P_{\mathrm{DE}}$
\\[0mm]
\hline
$q \geq 1$ [Type I (``Big Rip'') singularity] 
& $H \to \infty$
& $\dot{H} \to \infty$
& $J_1 \neq 0$ \quad 
& $J_1 \neq 0$ 
%$J_1 \neq 0$ or $J_2 \neq 0$
\\[0mm]
&
&
&
& or $J_2 \neq 0$
\\[0mm]
$0 < q < 1$ [Type III singularity] 
& $H \to \infty$
& $\dot{H} \to \infty$
& $J_1 \neq 0$ \quad 
& $J_1 \neq 0$
\\[0mm]
$-1 < q < 0$ [Type II (``sudden'') singularity] \quad 
& $H \to H_{\mathrm{s}}$
& $\dot{H} \to \infty$
& 
& $J_2 \neq 0$
\\[0mm]
$q < -1$, but $q$ is not any integer 
%[Type IV singularity]
& $H \to H_{\mathrm{s}}$
& $\dot{H} \to 0$
& 
& 
\\[0mm]
[Type IV singularity]
& 
& (Higher 
%derivatives of $H$ diverge)
&
&
\\[0mm]
&
& derivatives
&
&
\\[0mm]
&
& of $H$ diverge.)
&
&
\\[1mm]
\hline
\hline
\end{tabular}
\end{center}
\label{tb:table1}
\end{table*}
%%%%%%%%%%%%%%%%%%%%%%%%%%%%%%%%%%%%

%%%%%%%%%%%%%%%%%%%
%%%  Sec. IV
%%%%%%%%%%%%%%%%%%%
\section{Reconstruction of $f(T)$ gravity}

In this section, first we reconstruct $f(T)$ gravity 
in which there appear the finite-time future singularities 
discussed in Sec.~III\footnote{
In Appendix, 
we also describe a reconstruction method of $f(T)$ gravity by 
way of using a scalar field through the extension of that of $f(R)$ 
gravity~\cite{Reconstruction-Method, 
%Nojiri:2006gh, Nojiri:2006be, Nojiri:2008fk, 
Nojiri:2009kx, Nojiri:2011kd}.}. 
Next, we examine a correction term removing the finite-time future 
singularities.

%%%%%%%%%%%%%%%%%%%
%%%  Sec. IV A
%%%%%%%%%%%%%%%%%%%
\subsection{Reconstruction}

By using Eqs.~(\ref{eq:4.3}) and (\ref{eq:4.4}), we find that 
the effective EoS 
for the universe at the dark energy dominated stage is written as
\begin{equation}
w_{\mathrm{eff}} \approx w_{\mathrm{DE}} 
= \frac{P_{\mathrm{DE}}}{\rho_{\mathrm{DE}}}
= \frac{-\left[ 
4\left(1 -F -2TF^{\prime} \right) \dot{H} 
+\left( -T -f +2TF \right)
\right]}{-T -f +2TF}\,. 
\label{eq:IVB-1} 
\end{equation}
As another description, from Eqs.~(\ref{eq:4.3}) and (\ref{eq:4.4}) 
we have 
\begin{equation} 
P_{\mathrm{DE}} = -\rho_{\mathrm{DE}} 
+ I(H, \dot{H})\,,
\label{eq:IVB-2}
\end{equation} 
where 
\begin{equation} 
I \equiv 
- \frac{1}{{\kappa}^2} 
\left[ 
2\left(1 -F -2TF^{\prime} \right) \dot{H} \right]\,. 
\label{eq:IVB-3}
\end{equation} 
Since $T = -6H^2$, the form of $f(T)$ is a function of $H$. 
It follows from Eqs.~(\ref{eq:III-add-01}) and (\ref{eq:III-add-02}) that 
$P_{\mathrm{eff}} = - \rho_{\mathrm{eff}} -2\dot{H}/\kappa^2$. 
By comparing this equation with Eq.~(\ref{eq:IVB-2}), we acquire 
the differential equation 
\begin{equation} 
\dot{H} + 
\frac{{\kappa}^2}{2} I(H, \dot{H}) = 0\,. 
\label{eq:IVB-4}
\end{equation} 
The substitution of Eq.~(\ref{eq:IVB-3}) into Eq.~(\ref{eq:IVB-4}) 
yields 
\begin{equation} 
\dot{H} \left( F +2TF^{\prime} \right) = 0\,. 
\label{eq:IVB-5}
\end{equation} 
For $H$ in Eq.~(\ref{eq:2.13}), Eq.~(\ref{eq:IVB-5}) reads 
$F +2TF^{\prime} = 0$ because $\dot{H} \neq 0$.

%%%%%%%%%%%%%%%%%%%
%%%  Sec. IV A 1
%%%%%%%%%%%%%%%%%%%
\subsubsection{Power-law model}

As a form of $f(T)$, first we take a power-law model, given by 
\begin{equation} 
f(T) = A T^{\alpha}\,, 
\label{eq:IVA2-addition-01}
\end{equation} 
where $A (\neq 0)$ and $\alpha (\neq 0)$ are non-zero constants. 
In this case, from Eq.~(\ref{eq:IVB-5}) we have 
\begin{equation} 
F +2TF^{\prime} 
= A\left(-6\right)^{\alpha -1} \left(2\alpha-1\right) 
H^{2\left(\alpha-1\right)} 
= 0\,.
\label{eq:FT4-6-001}
\end{equation} 
In the limit of $t \to t_{\mathrm{s}}$, Eq.~(\ref{eq:FT4-6-001}) 
has to be satisfied. 
{}From Eqs.~(\ref{eq:2.13}) and (\ref{eq:IIIB-add-001}), we find that 
for $q > 0$ (i.e., the Type I singularity [$q \geq 1$] and the Type III 
singularity [$0<q<1$]), 
$\alpha < 1$, so that Eq.~(\ref{eq:FT4-6-001}) can be satisfied 
asymptotically, 
whereas 
for $q < 0$ (i.e., the Type II singularity [$-1<q<0$] and 
the Type IV singularity [$q<-1$]), 
$\alpha = 1/2$, in which Eq.~(\ref{eq:FT4-6-001}) is always satisfied. 

Furthermore, we state the meaning of the condition 
$F +2TF^{\prime} = 0$, which follows from Eq.~(\ref{eq:IVB-5}), 
and another condition that the Friedmann equation (\ref{eq:4.1}) can be 
satisfied, which may be interpreted as a consistency condition. 
Equations (\ref{eq:4.1}) and (\ref{eq:4.2}) are expressed as 
\begin{eqnarray} 
T + J_1 \Eqn{=} 0\,, 
%-f +2TF = 0\,, 
\label{eq:IV-BB-add-01} \\ 
J_2 - \dot{H} \Eqn{=} 0\,, 
%\left(J_2 -1 \right) \dot{H} \Eqn{=} 0\,, 
%-F -2TF^{\prime} = 0\,,
\label{eq:IV-BB-add-02} 
\end{eqnarray}
which are rewritten to 
\begin{eqnarray} 
%T + J_1 \Eqn{=} 0\,, 
-f +2TF = 0\,, 
\label{eq:IV-BB-add-03} \\ 
%\left(J_2 -1 \right) \dot{H} \Eqn{=} 0\,, 
-F -2TF^{\prime} = 0\,. 
\label{eq:IV-BB-add-04} 
\end{eqnarray}
Here, we have used Eqs.~(\ref{eq:IIB-Add-01}) and (\ref{eq:IIB-Add-02}) and 
$\dot{H} \neq 0$ for $H$ in Eqs.~(\ref{eq:2.13}) and (\ref{eq:IIIB-add-001}). 
%%%%%
%%%%%
It is clearly seen that 
Eq.~(\ref{eq:IV-BB-add-03}) corresponds to a consistency condition 
and that the relation $F +2TF^{\prime} = 0$ shown above is equivalent to 
Eq.~(\ref{eq:IV-BB-add-04}). 
Thus, the first condition is to satisfy the second gravitational equation 
(\ref{eq:4.2}). 
Moreover, from Eq.~(\ref{eq:IV-BB-add-03}) with 
Eq.~(\ref{eq:IVA2-addition-01}) 
we find that the consistency condition becomes 
\begin{equation}
-f +2TF = 
A\left(-6\right)^{\alpha} \left(2\alpha-1\right) 
H^{2\alpha} 
= 0\,.
\label{eq:FT4-6-002}
\end{equation} 
In the limit of $t \to t_{\mathrm{s}}$, Eq.~(\ref{eq:FT4-6-002}) 
must be satisfied. 
For $q > 0$ (i.e., the Type I singularity [$q \geq 1$] and the Type III singularity [$0<q<1$]), 
$\alpha < 0$, so that Eq.~(\ref{eq:FT4-6-002}) can be satisfied 
asymptotically, 
whereas 
for $q < 0$ (i.e., the Type II singularity [$-1<q<0$] and 
the Type IV singularity [$q<-1$]), 
$\alpha = 1/2$, in which Eq.~(\ref{eq:FT4-6-002}) is always satisfied. 

In addition, 
by using Eqs.~(\ref{eq:IIB-Add-01}) and (\ref{eq:IIB-Add-02}) with 
(\ref{eq:IVA2-addition-01}), we obtain 
\begin{eqnarray} 
J_1 \Eqn{=} 6H^2 \left[ 1-A\left(-6\right)^{\alpha -1} \left(2\alpha-1\right) 
H^{2\left(\alpha-1\right)}
\right]\,, 
\label{eq:IVA2-addition-02} \\  
J_2 \Eqn{=} \dot{H} \left[ 1-A\left(-6\right)^{\alpha -1} \alpha 
\left(2\alpha-1\right) H^{2\left(\alpha-1\right)} 
\right]\,. 
\label{eq:IVA2-addition-03}
\end{eqnarray}
For the expression of $H$ in Eqs.~(\ref{eq:2.13}) and 
(\ref{eq:IIIB-add-001}), from Eqs.~(\ref{eq:IVA2-addition-02}) and 
(\ref{eq:IVA2-addition-03}) we find that 
the conditions on $J_1$ and $J_2$ 
for the finite-time future singularities to exist in Table \ref{tb:table1} 
is always satisfied. 
Thus, there can appear all the four types of the finite-time future 
singularities. 
We note that 
for $\alpha = 1/2$, 
Eqs.~(\ref{eq:IVA2-addition-02}) and (\ref{eq:IVA2-addition-03}) becomes 
$
J_1 = -T \neq 0
$
and 
$
J_2 = \dot{H} \neq 0
$, 
respectively. 
Moreover, if $A=1$ and $\alpha = 1$, this model in Eq.~(\ref{eq:IIB-Add-01}) 
is equivalent to general relativity. 

%%%%%%%%
It is very important to note that the conditions: 
[for $q > 0$ (i.e., the Type I singularity [$q \geq 1$] and the Type III singularity [$0<q<1$]), $\alpha < 0$, whereas for $q < 0$ (i.e., the Type II singularity [$-1<q<0$] and the Type IV singularity [$q<-1$]), $\alpha = 1/2$] 
derived in the above considerations are ``necessary conditions'' to produce the finite-time future singularities and not sufficient conditions. 
Indeed, if $\alpha < 0$, the Type I singularity with $q \geq 1$ rather than 
the Type III singularity with $0<q<1$ appears because
in the limit of $t \to t_{\mathrm{s}}$, 
both $H$ and $\dot{H}$ with $q \geq 1$ diverge more rapidly than 
those with $0<q<1$. This originates from the absolute value of the power $q$, 
namely, the absolute value of $q$ for the Type I singularity ($q \geq 1$) is 
larger than that for the Type III singularity ($0<q<1$). 
As a result, the Type I singularity is realized faster than the Type III singularity, and eventually the Type I singularity appears. 
Similarly,
if $\alpha = 1/2$, the Type IV singularity with $q<-1$ rather than 
the Type II singularity with $-1<q<0$ occurs because
in the limit of $t \to t_{\mathrm{s}}$, 
$H \to H_{\mathrm{s}}$ and $\dot{H} \to 0$ wiht $q<-1$ are realized more 
quickly than $H \to H_{\mathrm{s}}$ and $\dot{H} \to \infty$ with $-1<q<0$. 
This also comes from the absolute value of the power $q$, 
namely, the absolute value of $q$ for the Type IV singularity ($q<-1$) is 
larger than that for the Type II singularity ($-1<q<0$).  
As a consequence, the Type IV singularity is produced faster than the Type II singularity, and accordingly the Type IV singularity appears. 
%%%%%%%%

%%%%%%%%
We also remark that in the Type V (``$w$'') singularity, 
a scale factor can be taken as~\cite{Dabrowski:2009zzb} 
\begin{eqnarray} 
a (t) \Eqn{=} a_{\mathrm{s}} \left( 1-\frac{3\sigma}{2} 
\left\{ \frac{n-1}{n-\left[ 2/\left(3\sigma\right) \right]}
\right\}^{n-1}\right)^{-1} 
+ \frac{1-2/\left(3\sigma \right)}{n-2/\left(3\sigma \right)} 
\nonumber \\ 
&& 
\times 
{}n a_{\mathrm{s}} 
\left( 1-\frac{2}{3\sigma} 
\left\{ \frac{n-\left[ 2/\left(3\sigma\right) \right]}{n-1}
\right\}^{n-1}\right)^{-1} 
\left( \frac{t}{t_{\mathrm{s}}} 
\right)^{2/\left(3\sigma \right)} 
\nonumber \\ 
&& 
{}+a_{\mathrm{s}} \left( \frac{3\sigma}{2} 
\left\{ \frac{n-1}{n-\left[ 2/\left(3\sigma\right) \right]}
\right\}^{n-1} -1 \right)^{-1} 
\left[ 1
- \frac{1-2/\left(3\sigma \right)}{n-2/\left(3\sigma \right)} 
\frac{t}{t_{\mathrm{s}}} 
\right]^{n}
\,, 
\label{eq:FT4-13-Add-IIIB-01} 
\end{eqnarray}
where 
$\sigma$ and $n$ are arbitrary constants. 
In the limit of $t \to t_{\mathrm{s}}$, 
$H(t \to t_{\mathrm{s}}) \to 0$ and $\dot{H} (t \to t_{\mathrm{s}}) \to 0$. 
On the other hand, the effective EoS for the universe 
$w_{\mathrm{eff}} = \left(1/3\right) \left(2q_{\mathrm{dec}} -1\right) 
\to \infty$. Here, $q_{\mathrm{dec}} \equiv -\ddot{a}a/\dot{a}^2$ 
is the deceleration parameter, which will again be redefined in 
Eq.~(\ref{eq:FT4-8-IVC2-addition-02}) in Sec.~IV C 2. 
%%%%%%%%
Thus, in the limit of $t \to t_{\mathrm{s}}$, 
since $\dot{H} (t \to t_{\mathrm{s}}) = 0$, from Eq.~(\ref{eq:IVB-5}) 
we obtain $F +2TF^{\prime} = 0$. 
if we take a power-law model in Eq.~(\ref{eq:IVA2-addition-01})  
with $A \neq 0$ and $\alpha > 1$, 
Eq.~(\ref{eq:FT4-6-001}) can be satisfied asymptotically  
because  $\dot{H} (t \to t_{\mathrm{s}}) = 0$. 
As a result, 
if we take a power-law model in Eq.~(\ref{eq:IVA2-addition-01}) 
with $A \neq 0$ and $\alpha > 1$, 
the Type V (``$w$'') singularity can appear. 
%%%%%%%%

%%%%%%%%%%%%%%%%%%%
%%%  Sec. IV A 2
%%%%%%%%%%%%%%%%%%%
\subsubsection{Exponential model}

Next, we examine an exponential model
\begin{equation} 
f(T) = C \exp \left( \lambda T \right)\,, 
\label{eq:FT4-6-003}
\end{equation} 
where $C (\neq 0)$ and $\lambda (\neq 0)$ are non-zero constants. 
%$C (\neq 0)$ is a non-zero constant and $\lambda (> 0)$ 
%is a positive constant.
In this case, Eqs.~(\ref{eq:IV-BB-add-03}) and (\ref{eq:IV-BB-add-04}) 
becomes 
\begin{eqnarray} 
-f +2TF \Eqn{=} C\left( -1 + 2\lambda T  \right) = 0 \,, 
\label{eq:FT4-6-004} \\   
-F -2TF^{\prime} \Eqn{=} -C\lambda \left( 2\lambda T + 1 \right) 
\exp \left( \lambda T \right) = 0\,. 
\label{eq:FT4-6-005} 
\end{eqnarray}
For the expression of $H$ in Eqs.~(\ref{eq:2.13}) and (\ref{eq:IIIB-add-001}), 
in the limit of $t \to t_{\mathrm{s}}$, 
both Eqs.~(\ref{eq:FT4-6-004}) and (\ref{eq:FT4-6-005}) cannot be satisfied 
simultaneously. Thus, in an exponential model in Eq.~(\ref{eq:FT4-6-003}) 
there cannot apper the finite-time future singularities.

%%%%%%%%%%%%%%%%%%%
%%%  Sec. IV A 3
%%%%%%%%%%%%%%%%%%%
\subsubsection{Logarithmic model}

Next, we explore an logarithmic model
\begin{equation} 
f(T) = D \ln \left( \gamma T \right)\,, 
\label{eq:FT4-6-006}
\end{equation} 
where $D (\neq 0)$ is a non-zero constant and $\lambda (> 0)$ 
is a positive constant. 
In this case, Eqs.~(\ref{eq:IV-BB-add-03}) and (\ref{eq:IV-BB-add-04}) 
becomes 
\begin{eqnarray} 
-f +2TF \Eqn{=} D\left[ - \ln \left( \gamma T \right) + 2 \right) = 0 \,, 
\label{eq:FT4-6-007} \\   
-F -2TF^{\prime} \Eqn{=} \frac{D}{T} = 0\,. 
\label{eq:FT4-6-008} 
\end{eqnarray}
For the expression of $H$ in Eqs.~(\ref{eq:2.13}) and (\ref{eq:IIIB-add-001}), 
in the limit of $t \to t_{\mathrm{s}}$, 
both Eqs.~(\ref{eq:FT4-6-007}) and (\ref{eq:FT4-6-008}) cannot be satisfied 
simultaneously. 
Hence, in an logarithmic model in Eq.~(\ref{eq:FT4-6-006}) 
the finite-time future singularities cannot occur, 
similarly to the case of a exponential model 
in Eq.~(\ref{eq:FT4-6-003}) in Sec.~IV A 2. 
%%%%%
Thus, in general the occurrence of the finite-time future singularities 
in $f(T)$ gravity is realized in less cases than in $f(R)$ gravity. 
%%%%%

%%%%%%%%%%%%%%%%%%%
%%%  Sec. IV B
%%%%%%%%%%%%%%%%%%%
\subsection{Correction term removing the finite-time future singularities}

It is known that in $f(R)$ gravity, 
the addition of an $R^2$ term can cure the finite-time future singularities 
(see Ref.~\cite{Review-Nojiri-Odintsov}). 
Recently, it has also been demonstrated in Ref.~\cite{Bamba:2011ky} that 
the addition of an $R^2$ term can remove the finite-time future singularities 
in non-local gravity. 
%%%%%%%%
In this subsection, 
we investigate a correction term for the form of $f(T)$ 
in Eq.~(\ref{eq:IVA2-addition-01}) so that the finite-time future 
singularities cannot appear. 
To execute this analysis, we explore an additional term of 
as function of $T$ to the form of $f(T)$ 
in Eq.~(\ref{eq:IVA2-addition-01}) so that 
for $H$ in Eqs.~(\ref{eq:2.13}) and (\ref{eq:IIIB-add-001}), 
the gravitational field equations (\ref{eq:4.1}) 
and (\ref{eq:4.2}) with Eqs.~(\ref{eq:4.3}) and (\ref{eq:4.4}) 
cannot be satisfied. 
%%%%%%%%

As a straight forward procedure, 
we explore the case that 
the form of $f(T)$ represented by Eq.~(\ref{eq:IVA2-addition-01}) 
has a correction term $f_{\mathrm{c}} (T)$, 
and analyze whether 
Eqs.~(\ref{eq:IV-BB-add-03}) and (\ref{eq:IV-BB-add-04}) 
can be satisfied or not. 
As an example, we choose a correction term $f_{\mathrm{c}} (T)$ as 
\begin{equation} 
f_{\mathrm{c}} (T) = B T^\beta\,, 
\label{eq:IV-BB-add-05}
\end{equation} 
where $B (\neq 0)$ and $\beta (\neq 0)$ are non-zero constants. 
For $\beta = 2$, the correction term is similarly to that in $f(R)$ gravity, 
i.e., a $T^2$ term. 
By combining Eqs.~(\ref{eq:IVA2-addition-01}) and (\ref{eq:IV-BB-add-05}), 
the total form of $f(T)$ including the correction term is expressed as 
\begin{equation} 
f(T) = A T^{\alpha} + B T^\beta\,. 
\label{eq:IV-BB-add-06}
\end{equation} 
By substituting Eq.~(\ref{eq:IV-BB-add-06}) into Eqs.~(\ref{eq:IV-BB-add-03}) 
and (\ref{eq:IV-BB-add-04}), we find 
\begin{eqnarray}  
-f +2TF \Eqn{=} A\left(2\alpha-1\right) T^{\alpha} 
+ B\left(2\beta-1\right) T^{\beta} \neq 0\,, 
\label{eq:IV-BB-add-07} \\  
-F -2TF^{\prime} \Eqn{=} -A\alpha\left(2\alpha-1\right) T^{\alpha-1} 
-B\beta\left(2\beta-1\right) T^{\beta-1} \neq 0\,. 
\label{eq:IV-BB-add-08} 
\end{eqnarray}
{}From the considerations in Sec. IV A, we find that 
for $q > 0$ (i.e., the Type I singularity [$q \geq 1$] and the Type III singularity [$0<q<1$]), 
$\beta > 0$, so that the second inequality in 
Eq.~(\ref{eq:IV-BB-add-07}) can be satisfied 
asymptotically, 
whereas 
for $q < 0$ (i.e., the Type II singularity [$-1<q<0$] and 
the Type IV singularity [$q<-1$]), 
$\beta \neq 1/2$, in which the second inequality in 
Eq.~(\ref{eq:IV-BB-add-08}) is always satisfied. 
As a result, 
if $\beta >1$, for the Hubble parameter in Eqs.~(\ref{eq:2.13}) 
and (\ref{eq:IIIB-add-001}), 
in the limit of $t \to t_{\mathrm{s}}$ 
both of the gravitational field equations 
(\ref{eq:4.1}) and (\ref{eq:4.2}) cannot be satisfied. 
This means that 
a power-low type correction term $T^\beta$ with $\beta>1$ 
can remove the finite-time future singularities in $f(T)$ gravity. 
%%%%%
In Table~\ref{tb:table2}, we describe necessary conditions on the model 
parameters of a power-law model of 
$f(T)$ in Eq.~(\ref{eq:IVA2-addition-01}) with realizing the finite-time 
future singularities, 
the emergence of the finite-time future singularities, and 
those of the correction term 
$f_{\mathrm{c}} (T) = B T^\beta$ in Eq.~(\ref{eq:IV-BB-add-05}) with removing 
the finite-time future singularities. 
%%%%%
It is interesting to emphasize that a $T^2$ term, i.e., $\beta = 2$, 
which is the minimum integer to satisfy the condition $\beta>1$, 
can remove all the four types of the finite-time future singularities 
in $f(T)$ gravity. This consequence is the same as that in $f(R)$ gravity. 

%%%%%%%%
It is interesting note that for the case of the Type V (``$w$'') singularity, 
since in the limit of $t \to t_{\mathrm{s}}$ 
$H (t \to t_{\mathrm{s}}) = 0$, 
if we choose a power-law type correction term $f_{\mathrm{c}} (T)$ in 
Eq.~(\ref{eq:IV-BB-add-05}) with $B \neq 0$ and $\beta<0$ 
and substitute Eq.~(\ref{eq:IV-BB-add-06}) into Eqs.~(\ref{eq:IV-BB-add-03}) 
and (\ref{eq:IV-BB-add-04}), we find 
Eqs.~(\ref{eq:IV-BB-add-07}) and 
(\ref{eq:IV-BB-add-08}). In other words, 
both of the gravitational field equations 
(\ref{eq:4.1}) and (\ref{eq:4.2}) cannot be satisfied asymptotically. 
This means that 
a power-law type correction term $f_{\mathrm{c}} (T)$ in 
Eq.~(\ref{eq:IV-BB-add-05}) with $B \neq 0$ and $\beta<0$ 
can remove the Type V (``$w$'') singularity.
%%%%%%%%

%%%%%%%%% Table II %%%%%%%%%%%
\begin{table*}[tbp]
\caption{Necessary conditions on the model parameters of a power-law model of 
$f(T)$ in Eq.~(\ref{eq:IVA2-addition-01}) with realizing the finite-time 
future singularities, the emergence of the finite-time future singularities, 
and those of the correction term 
$f_{\mathrm{c}} (T) = B T^\beta$ in Eq.~(\ref{eq:IV-BB-add-05}) with removing 
the finite-time future singularities.}
\begin{center}
\begin{tabular}
{lcll}
\hline
\hline
$q (\neq 0, \, -1)$
& Emergence\,\,\,\,\, 
& $f(T) = A T^\alpha$ 
& $f_{\mathrm{c}} (T) = B T^\beta$ 
\\[0mm]
&
& ($A \neq 0$, $\alpha \neq 0$)\,\,\,\,\,
& ($B \neq 0$, $\beta \neq 0$)
\\[0mm]
\hline
$q \geq 1$ [Type I (``Big Rip'') singularity] 
& Yes 
& $\alpha < 0$
& $\beta > 1$
\\[0mm]
$0 < q < 1$ [Type III singularity] 
& ---
& $\alpha < 0$
& $\beta > 1$
\\[0mm]
$-1 < q < 0$ [Type II (``sudden'') singularity]
& ---
& $\alpha = 1/2$
& $\beta \neq 1/2$
\\[0mm]
$q < -1$, but $q$ is not any integer 
%[Type IV singularity]
\,\,\,\,\,
& Yes
& $\alpha = 1/2$
& $\beta \neq 1/2$
\\[0mm]
[Type IV singularity]
&
&
\\[1mm]
\hline
\hline
\end{tabular} 
\end{center}
\label{tb:table2}
\end{table*}
%%%%%%%%%%%%%%%%%%%%%%%%%%%%%%%%%%%%

%%%%%%%%%%%%%%%%%%%
%%%  Sec. IV C
%%%%%%%%%%%%%%%%%%%
\subsection{Reconstructed models realizing cosmologies}

In this subsection, we explicitly present the reconstruction of $f(T)$ models 
with realizing (a) inflation in the early universe, (b) 
the $\Lambda$CDM model, (c) Little Rip cosmology and (d) Pseudo-Rip cosmology.

%%%%%%%%%%%%%%%%%%%
%%%  Sec. IV C 1
%%%%%%%%%%%%%%%%%%%
\subsubsection{Inflation in the early universe}

For generality, we consider power-law inflation. 
We suppose that the Hubble parameter is given by 
\begin{equation}
H = \frac{h_{\mathrm{inf}}}{t}\,, 
\label{eq:IVC1-ADD-001}
\end{equation}
where $h_{\mathrm{inf}} (> 1)$ is a constant larger than unity. 
%where $h_{\mathrm{inf}} (> 0)$ is a non-zero constant. 
It follows from Eq.~(\ref{eq:IVC1-ADD-001}) that the scale factor is given by 
\begin{equation}
a(t) = a_{\mathrm{inf}} t^{h_{\mathrm{inf}}}\,.
\label{eq:IVC1-ADD-002}
\end{equation}
%
%where $q (>1)$ is a constant larger than unity. 
{}From Eq.~(\ref{eq:IVC1-ADD-002}), we find that 
$\ddot{a} = a_{\mathrm{inf}} h_{\mathrm{inf}} 
\left( h_{\mathrm{inf}} - 1 \right)t^{h_{\mathrm{inf}}-2} > 0$, 
and hence power-law inflation occurs. 
In this case, since $\dot{H} = -h_{\mathrm{inf}}/t^2 \neq 0$, by using 
Eq.~(\ref{eq:IVB-5}) we obtain the condition $F +2TF^{\prime} = 0$. 

As explained in Sec.~IV A 1, for a power-law model $f(T) = A T^{\alpha}$ 
in Eq.~(\ref{eq:IVA2-addition-01}), 
the conditions to be satisfied, which originate from the 
gravitational equations (\ref{eq:4.1}) and (\ref{eq:4.2}), 
are given by Eqs.~(\ref{eq:FT4-6-001}) and (\ref{eq:FT4-6-002}). 
Therefore, if $\alpha <0$, 
in the limit of $t \to 0$, 
$H$ in Eq.~(\ref{eq:IVC1-ADD-001}) diverges, so that 
Eqs.~(\ref{eq:FT4-6-001}) and (\ref{eq:FT4-6-002}) can approximately be 
satisfied in the very early universe. 
Moreover, if $\alpha = 1/2$, Eqs.~(\ref{eq:FT4-6-001}) and 
(\ref{eq:FT4-6-002}) can always be met.

%%%%%%%%%%%%%%%%%%%
%%%  Sec. IV C 2
%%%%%%%%%%%%%%%%%%%
\subsubsection{The $\Lambda$CDM model}

When we describe the $\Lambda$CDM model by the action 
in Eq.~(\ref{eq:2.6}), 
$f(T) = T - 2\Lambda$, where $\Lambda >0$ is positive cosmological constant, as is in general relativity. In this case, by substituting this form of $f(T)$ 
into Eqs.~(\ref{eq:4.1}) and (\ref{eq:4.2}), we have 
\begin{eqnarray}
H^2 \Eqn{=} \frac{\Lambda}{3}\,, 
\label{eq:FT4-6-IVC2-01} \\ 
\dot{H}
\Eqn{=} 0\,. 
\label{eq:FT4-6-IVC2-02} 
\end{eqnarray}
Clearly, from Eqs.~(\ref{eq:FT4-6-IVC2-01}) and (\ref{eq:FT4-6-IVC2-02}), 
we find $H \equiv H_{\Lambda} = \sqrt{\Lambda/3} = \mathrm{constant}$, 
where we have defined the Hubble parameter at the cosmological constant 
dominated stage as $H_{\Lambda} (>0)$. 
%and therefore the late-time cosmic acceleration can occur. 
%In the $\Lambda$CDM model, 
Furthermore, 
%by using Eqs.~(\ref{eq:FT4-6-IVC2-01})
the scale factor is expressed as 
\begin{equation}
a = a_{\Lambda} 
\exp \left( H_{\Lambda} t \right)\,, 
\label{eq:FT4-8-IVC2-addition-01} 
\end{equation}
where $a_{\Lambda} (>0)$ 
%and $H_{\Lambda} (>0)$ are 
is a positive constant. 

In the $\Lambda$CDM model, from Eq.~(\ref{eq:2.17}) we find that 
the EoS is given by $w_{\mathrm{DE}} = -1$ due to the fact that 
$H$ is constant. 
Moreover, 
the deceleration parameter $q_{\mathrm{dec}}$, 
the jerk parameter $j$ and 
the snark parameter $s$ are defined by~\cite{Sahni:2002fz}
%Astashenok:2012tv} 
%[the second reference in Ref.~\cite{}] 
%
\begin{eqnarray}
q_{\mathrm{dec}} \Eqn{\equiv} -\frac{1}{aH^2} \frac{d^2 a}{dt^2}\,, 
\label{eq:FT4-8-IVC2-addition-02} \\ 
j \Eqn{\equiv} 
\frac{1}{aH^3} \frac{d^3 a}{dt^3}\,, 
\label{eq:FT4-8-IVC2-addition-03} \\ 
s \Eqn{\equiv} \frac{j - 1}{3 \left( q_{\mathrm{dec}} -1/2 \right)}\,.
\label{eq:FT4-8-IVC2-addition-04} 
\end{eqnarray}
By using Eq.~(\ref{eq:FT4-8-IVC2-addition-01}), 
we obtain $q_{\mathrm{dec}} = -1$, $j = 1$ and $s = 0$. 

%%%%%
The limit on a constant EoS for dark energy 
in a flat universe has been estimated as 
$w_{\mathrm{DE}} = -1.10 \pm 0.14 \, 
(68 \% \, \mathrm{CL})$ [the second reference in Ref.~\cite{Komatsu-WMAP}]. 
In addition, 
for a time-dependent EoS for dark energy 
with a linear form $w_{\mathrm{DE}}(a) = w_{\mathrm{DE}\,0} + 
w_{\mathrm{DE}\,a} \left( 1-a \right)$~\cite{C-P-L}, 
where $w_{\mathrm{DE}\,0}$ and $w_{\mathrm{DE}\,a}$ 
are the current value of $w_{\mathrm{DE}}$ and 
its derivative, respectively, 
the constraints have been analyzed as 
$w_{\mathrm{DE}\,0} = -0.93 \pm 0.13$ and 
$w_{\mathrm{DE}\,a} = -0.41^{+0.72}_{-0.71} \, (68 \% \, \mathrm{CL})$ 
[the second reference in Ref.~\cite{Komatsu-WMAP}]. 
%%%%%
Thus, 
the deviations of the values of $(w_{\mathrm{DE}}, q_{\mathrm{dec}}, j, s)$ 
from those for the $\Lambda$CDM model $(-1, -1, 1, 0)$ show how the model is 
different form the $\Lambda$CDM model. In other words, we can use 
these four parameters as a observational test. 
%%%%%

We remark that if we consider the early universe, the model 
$f(T) = T - 2\Lambda$ can lead to exponential inflation 
realizing de Sitter expansion of the universe, i.e., 
the Hubble parameter is given by 
\begin{equation}
H = H_{\mathrm{inf}} = \mathrm{constant}\,,
\label{eq:IVC1-ADD-003}
\end{equation}
where $H_{\mathrm{inf}} > 0$. 
By using Eq.~(\ref{eq:IVC1-ADD-003}), we have 
\begin{equation}
a(t) = a_{\mathrm{inf}} \exp \left( H_{\mathrm{inf}} t \right)\,, 
\label{eq:IVC1-ADD-004}
\end{equation}
where $a_{\mathrm{inf}} (> 0)$ is a positive constant.

%%%%%%%%%%%%%%%%%%%
%%%  Sec. IV C 3
%%%%%%%%%%%%%%%%%%%
\subsubsection{Little Rip cosmology}

Furthermore, 
we study Little Rip cosmology~\cite{Frampton:2011sp, Brevik:2011mm, 
%FLNOS-ANOY, 
Frampton:2011rh, Astashenok:2012tv, 
Nojiri:2011kd, GL-IT, Ito:2011ae, XZL-MKO}, 
which corresponds to a mild phantom scenario. 
The Little Rip scenario has been proposed to avoid 
the finite-time future singularities, in particular a Big Rip singularity. 
In this scenario, the energy density of dark energy increases in time with 
$w_{\mathrm{DE}}$ being less than $-1$ and then $w_{\mathrm{DE}}$ 
asymptotically approaches $w_{\mathrm{DE}} = -1$. 
However, such a scenario eventually leads to the dissolution of 
bound structures at some time in the future via the increase of an inertial 
force between objects. This process is called 
the ``Little Rip''. 

As an example to realize Little Rip cosmology, we take 
the Hubble parameter as~\cite{Frampton:2011rh} 
\begin{equation}
H = H_{\mathrm{LR}} \exp \left( \xi t \right)\,, 
\label{eq:FT4-6-IVC3-01}
\end{equation}
where $H_{\mathrm{LR}} (>0)$ and $\xi (>0)$ are positive constants. 
In this case, the scale factor $a$ is expressed as 
\begin{equation}
a = a_{\mathrm{LR}} \exp \left[ \frac{H_{\mathrm{LR}}}{\xi} 
\exp \left( \xi t \right)
\right]\,, 
\label{eq:FT4-8-IVC3-ADDING-1}
\end{equation}
where $a_{\mathrm{LR}} (>0)$ is a positive constant. 
Moreover, from Eq.~(\ref{eq:2.17}) we obtain 
\begin{equation}
w_{\mathrm{DE}} = -1 - \frac{2 \xi}{3 H_{\mathrm{LR}}} 
\exp \left( -\xi t \right)\,. 
\label{eq:FT4-6-IVC3-wDE}
\end{equation}
Since $\dot{H} = H_{\mathrm{LR}} \xi \exp \left( \xi t \right) > 0$, 
$w_{\mathrm{DE}} < -1$, i.e., the universe is always in the phantom phase. 
In the limit of $t \to \infty$, we find $w_{\mathrm{DE}} \to -1$ and 
hence the Little Rip scenario can be realized.

%%%%%%%%
In the expression of $w_{\mathrm{DE}}$ in Eq.~(\ref{eq:FT4-6-IVC3-wDE}), 
if we take $\xi = H_0$, at $t = t_0 \approx H_0^{-1}$, 
we have $w_{\mathrm{DE}} = -1-2H_0/\left(3H_{\mathrm{LR}} e \right)$. 
%where $H_0$ is the current value of the Hubble parameter, 
Here, $t_0$ is the present time, $H_0$ is  
the current value of the Hubble parameter given by  
$H_{0} = 2.1 h \times 10^{-42} \, \mathrm{GeV}$~\cite{Kolb and Turner}
with $h = 0.7$ [the second reference in Ref.~\cite{Komatsu-WMAP}, 
\cite{Freedman:2000cf}], and $e = 2.71828$. 
By comparing this expression 
with the observational constraint on 
$w_{\mathrm{DE}} = -1.10 \pm 0.14 \, 
(68 \% \, \mathrm{CL})$ [the second reference in Ref.~\cite{Komatsu-WMAP}], 
we find that if $H_{\mathrm{LR}} \geq \left[2H_0/\left(3e\right)\right]
\left(1/0.24\right) = 1.50 \times 10^{-42} \, \mathrm{GeV}$, 
the current value of $w_{\mathrm{DE}}$ in this Little Rip model is consistent 
with the observations. 
Here, we have used the fact that $\xi = H_0$ and $H_{\mathrm{LR}}$ are 
positive values. 
%%%%%%%%

%%%%%
By using 
Eqs.~(\ref{eq:FT4-8-IVC2-addition-02})--(\ref{eq:FT4-8-IVC2-addition-04}), 
(\ref{eq:FT4-6-IVC3-01}) and (\ref{eq:FT4-8-IVC3-ADDING-1}), 
we acquire 
\begin{eqnarray}
q_{\mathrm{dec}} \Eqn{=} -1 -\frac{\xi}{H_{\mathrm{LR}} 
\exp \left( \xi t \right)}\,,
%-\frac{1}{aH^2} \frac{d^2 a}{dt^2}\,, 
\label{eq:FT4-8-IVC3-addition-01} \\ 
j \Eqn{=} 1 + \frac{\xi}{H_{\mathrm{LR}}} 
\left[ \frac{\xi}{H_{\mathrm{LR}} \exp \left( \xi t \right)} + 3 \right] 
\frac{1}{\exp \left( \xi t \right)}\,, 
%-\frac{1}{aH^3} \frac{d^3 a}{dt^3}\,, 
\label{eq:FT4-8-IVC3-addition-02} \\ 
s \Eqn{=} -\frac{2\xi \left[ \xi + 3 H_{\mathrm{LR}} \exp \left( \xi t \right) \right]}{3H_{\mathrm{LR}} 
\left[ 2\xi + 3 H_{\mathrm{LR}} \exp \left( \xi t \right) \right] 
\exp \left( \xi t \right)}\,. 
%\frac{j - 1}{3 \left( q_{\mathrm{dec}} -1/2 \right)}\,.
\label{eq:FT4-8-IVC3-addition-03} 
\end{eqnarray}
% 
%%%%%
%%%%%%%%
For $\xi = H_0$, at $t = t_0 \approx H_0^{-1}$, from 
Eqs.~(\ref{eq:FT4-6-IVC3-wDE})--(\ref{eq:FT4-8-IVC3-addition-03}), 
we describe the expression of $w_{\mathrm{DE}}$, 
$q_{\mathrm{dec}}$, $j$ and $s$ at the present time $t_0$ as 
\begin{eqnarray}
w_{\mathrm{DE}} (t=t_0) \Eqn{=} -1 -\frac{2}{3}\chi\,,
\label{eq:FT4-10-IVC3-ADDING-01} \\ 
q_{\mathrm{dec}} (t=t_0) \Eqn{=} -1 - \chi\,,
\label{eq:FT4-10-IVC3-ADDING-02} \\ 
j (t=t_0) \Eqn{=} 1 +  \chi \left( \chi + 3 \right)\,, 
\label{eq:FT4-10-IVC3-ADDING-03} \\ 
s (t=t_0) \Eqn{=} -\frac{2\chi \left( \chi + 3 \right)}{3 \left( 2\chi + 3 \right)}\,, 
\label{eq:FT4-10-IVC3-ADDING-04} 
\end{eqnarray}
with 
\begin{equation}
\chi \equiv \frac{H_0}{H_{\mathrm{LR}} e} \leq 0.36\,,
\label{eq:FT4-10-IVC3-ADDING-05}
\end{equation}
where the second inequality in Eq.~(\ref{eq:FT4-10-IVC3-ADDING-05}) follows 
from the observational constraint on 
$w_{\mathrm{DE}} = -1.10 \pm 0.14 \, 
(68 \% \, \mathrm{CL})$ [the second reference in Ref.~\cite{Komatsu-WMAP}] 
as $\chi \leq \left(3/2\right) 0.24 = 0.36$. 
As a result, if we take $\chi \ll 1$ enough for the deviation of 
the values of the four parameters $(w_{\mathrm{DE}}, q_{\mathrm{dec}}, j, s)$ 
from those for the $\Lambda$CDM model $(-1, -1, 1, 0)$ 
to be very small, this Little Rip model can be compatible 
with the $\Lambda$CDM model. 
%%%%%%%%

It follows from Eq.~(\ref{eq:FT4-6-IVC3-01}) that 
in the limit of $t \to \infty$, $H$ diverges. 
For a power-law model $f(T) = A T^{\alpha}$ 
in Eq.~(\ref{eq:IVA2-addition-01}), 
if $\alpha <0$, 
in the limit of $t \to \infty$, 
Eqs.~(\ref{eq:FT4-6-001}) and (\ref{eq:FT4-6-002}) can be 
satisfied asymptotically. 
This is just an opposite case in the model with realizing 
inflation in Sec.~IV C 1 because the limit in terms of $t$ is 
the opposite direction. 
In addition, 
if $\alpha = 1/2$, Eqs.~(\ref{eq:FT4-6-001}) and 
(\ref{eq:FT4-6-002}) can always be met, similarly to that 
in the case of inflation in Sec.~IV C 1.

%%%%%%%%%%%%%%%%%%%
%%%  Sec. IV C 4
%%%%%%%%%%%%%%%%%%%
\subsubsection{Pseudo-Rip cosmology}

We also investigate Pseudo-Rip cosmology~\cite{Astashenok:2012tv, 
Frampton:2011aa}. 
The above four cosmological models can be classified by using 
the behavior of the Hubble parameter as follows~\cite{Frampton:2011aa}. 
(a) power-law inflation: 
\begin{equation}
H(t) 
%\longrightarrow 
\to \infty\,, 
\quad 
t \to 0\,.
\label{eq:FT4-6-IVC4-add-001}
\end{equation}
(b) the $\Lambda$CDM model or exponential inflation: 
\begin{equation}
H(t) = H(t_0)\,.
\label{eq:FT4-6-IVC4-add-002}
\end{equation}
(c) Little Rip cosmology: 
\begin{equation}
H(t) 
%\longrightarrow 
\to\infty\,, 
\quad 
t \to \infty\,.
\label{eq:FT4-6-IVC4-add-003}
\end{equation}
(d) Pseudo-Rip cosmology, which is also phantom asymptotically de Sitter 
universe: 
\begin{equation}
H(t) 
%\longrightarrow 
\to H_\infty < \infty\,, 
\quad 
t \to \infty\,.
\label{eq:FT4-6-IVC4-add-004}
\end{equation}
Here, 
%$t_0$ is the present time and 
we consider $t \geq t_0$. 
Moreover, $H_\infty (>0)$ is a positive constant. 
We also note that for a Big Rip singularity, 
$H(t) 
%\longrightarrow 
\to \infty,\,\,\, t \to t_{\mathrm{s}}$, 
as shown in Table~\ref{tb:table1}. 
%%%%%%%%
As an example of a Pseudo-Rip model, we take 
\begin{equation}
H(t) = H_{\mathrm{PR}} \tanh \left( \frac{t}{t_0} \right)\,, 
\label{eq:FT4-6-IVC4-add-005}
\end{equation}
where $H_{\mathrm{PR}} (>0)$ is a positive constant. 
In this case, the scale factor $a$ is expressed as 
\begin{equation}
a = a_{\mathrm{PR}} \cosh \left( \frac{t}{t_0} \right)\,, 
\label{eq:FT4-8-IVC4-2nd-adding-01}
\end{equation}
where $a_{\mathrm{PR}} (>0)$ is a positive constant. 
{}From Eq.~(\ref{eq:FT4-6-IVC4-add-005}), 
we find that $H(t)$ is monotonically increasing function of $t$ and 
$H(t) 
%\longrightarrow 
\to H_{\mathrm{PR}} < \infty,\,\,\, t \to \infty$. 
Thus, a behavior of $H$ in the Pseudo-Rip cosmology in 
(\ref{eq:FT4-6-IVC4-add-004}) is realized. We also have 
$\dot{H}(t) = H_{\mathrm{PR}}/\left[ t_0 \cosh^2 \left( t/t_0 \right) \right] 
%\longrightarrow 
\to 0,\,\,\, t \to \infty$. This means from Eq.~(\ref{eq:4.2}) that 
$P \to -\rho$ in the limit of 
$t \to \infty$. 
For a power-law model $f(T) = A T^{\alpha}$ with $\alpha = 1/2$
in Eq.~(\ref{eq:IVA2-addition-01}), i.e., $f(T) = A \sqrt{T}$, 
Eqs.~(\ref{eq:4.2}) and (\ref{eq:4.2}) can always be satisfied 
including in the limit of $t \to \infty$.

%%%%%
For $H$ in Eq.~(\ref{eq:FT4-6-IVC4-add-005}), 
from Eq.~(\ref{eq:2.17}) we find that the EoS is given by 
\begin{equation}
w_{\mathrm{DE}} = 
-1 -\frac{2}{3 t_0 H_{\mathrm{PR}}} \frac{1}{\sinh^2 \left( t/t_0 \right)}\,.
\label{eq:FT4-8-IVC4-addition-01} 
\end{equation}
{}From Eq.~(\ref{eq:FT4-8-IVC4-addition-01}), we see that 
$w_{\mathrm{DE}} < -1$, namely, 
the universe is always in the phantom phase, 
because $\dot{H}(t) = 
H_{\mathrm{PR}}/\left[ t_0 \cosh^2 \left( t/t_0 \right) \right] > 0$. 
%$w_{\mathrm{DE}} < -1$, i.e., the universe is always in the phantom phase. 
In the limit of $t \to \infty$, we find $w_{\mathrm{DE}} \to -1$, similarly to 
that in Little Rip cosmology. 

%%%%%%%%
It follows from $w_{\mathrm{DE}}$ in Eq.~(\ref{eq:FT4-8-IVC4-addition-01}) 
that at $t = t_0 \approx H_0^{-1}$, 
$w_{\mathrm{DE}} = -1-\left[2H_0/\left(3H_{\mathrm{PR}}\right)\right] 
\left[4/\left(e - e^{-1}\right)^2\right]$. 
In comparison with the observational constraint on 
$w_{\mathrm{DE}} = -1.10 \pm 0.14 \, 
(68 \% \, \mathrm{CL})$ [the second reference in Ref.~\cite{Komatsu-WMAP}], 
we find that if $H_{\mathrm{PR}} \geq \left(2H_0/3\right) 
\left[4/\left(e - e^{-1}\right)^2\right] \left(1/0.24\right) = 2.96 
\times 10^{-42} \, \mathrm{GeV}$, 
the current value of $w_{\mathrm{DE}}$ in this Pseudo-Rip model is compatible 
with the observations. 
Here, we have used the fact that $t_0 \approx H_0^{-1}$ and $H_{\mathrm{PR}}$ 
are positive values. 
%%%%%%%%

Moreover, by using 
Eqs.~(\ref{eq:FT4-8-IVC2-addition-02})--(\ref{eq:FT4-8-IVC2-addition-04}), 
(\ref{eq:FT4-6-IVC4-add-005}) and (\ref{eq:FT4-8-IVC4-2nd-adding-01}), 
we have 
\begin{eqnarray}
%w_{\mathrm{DE}} \Eqn{=} 
%-1 -\frac{2}{3 t_0 H_{\mathrm{PR}}} \frac{1}{\sinh^2 \left( t/t_0 \right)}\,, 
%\label{eq:FT4-8-IVC4-addition-01} \\ 
q_{\mathrm{dec}} \Eqn{=} -1 + \frac{\left(t_0 H_{\mathrm{PR}}\right)^2 
\tanh^2 \left( t/t_0 \right) - 1}{ \left(t_0 H_{\mathrm{PR}}\right)^2 
\tanh^2 \left( t/t_0 \right)}\,,
%-\frac{1}{aH^2} \frac{d^2 a}{dt^2}\,, 
\label{eq:FT4-8-IVC4-addition-02} \\ 
j \Eqn{=} 1 + \frac{1-\left(t_0 H_{\mathrm{PR}}\right)^3 
\tanh^2 \left( t/t_0 \right)}{ \left(t_0 H_{\mathrm{PR}}\right)^3 
\tanh^2 \left( t/t_0 \right)}\,, 
%-\frac{1}{aH^3} \frac{d^3 a}{dt^3}\,, 
\label{eq:FT4-8-IVC4-addition-03} \\ 
s \Eqn{=} \frac{2}{3 t_0 H_{\mathrm{PR}}} \frac{
\left(t_0 H_{\mathrm{PR}}\right)^3 \tanh^2 \left( t/t_0 \right) -1}{
\left(t_0 H_{\mathrm{PR}}\right)^2 \tanh^2 \left( t/t_0 \right) +2}\,.
%\frac{j - 1}{3 \left( q_{\mathrm{dec}} -1/2 \right)}\,.
\label{eq:FT4-8-IVC4-addition-04} 
\end{eqnarray}
% 
%%%%%

%%%%%%%%
At $t = t_0 \approx H_0^{-1}$, from 
Eqs.~(\ref{eq:FT4-8-IVC4-addition-01})--(\ref{eq:FT4-8-IVC4-addition-04}), 
we describe the expression of $w_{\mathrm{DE}}$, 
$q_{\mathrm{dec}}$, $j$ and $s$ at the present time $t_0$ as 
\begin{eqnarray}
w_{\mathrm{DE}} (t=t_0) \Eqn{=} -1-\frac{2\delta}{3\sinh^2 1}\,,
\label{eq:FT4-10-IVC4-ADDING-01} \\ 
q_{\mathrm{dec}} (t=t_0) \Eqn{=} -1 + \frac{\delta^2 \tanh^2 1 - 1}{\delta^2 
\tanh^2 1}\,, 
\label{eq:FT4-10-IVC4-ADDING-02} \\ 
j (t=t_0) \Eqn{=} 1 + \frac{1-\delta^3 \tanh^2 1}{\delta^3 \tanh^2 1}\,, 
\label{eq:FT4-10-IVC4-ADDING-03} \\ 
s (t=t_0) \Eqn{=} \frac{2}{3 \delta} \frac{\delta^3 \tanh^2 1 -1}{\delta^2 
\tanh^2 1 +2}\,, 
\label{eq:FT4-10-IVC4-ADDING-04} 
\end{eqnarray}
with 
\begin{equation}
\delta \equiv \frac{H_0}{H_{\mathrm{PR}}} \leq 0.497196\,,
\label{eq:FT4-10-IVC4-ADDING-05}
\end{equation}
where the second inequality in Eq.~(\ref{eq:FT4-10-IVC4-ADDING-05}) follows 
from the observational constraint on 
$w_{\mathrm{DE}} = -1.10 \pm 0.14 \, 
(68 \% \, \mathrm{CL})$ [the second reference in Ref.~\cite{Komatsu-WMAP}] 
as $\delta \leq \left(3/2\right) 0.24 \sinh^2 1 = 0.497196$. 
Here, we use $\sinh^2 1 = 1.3811$ and $\tanh^2 1 = 0.580026$. 
As a consequence, 
we can take an appropriate value of $\delta$ in order for the deviation of 
the values of the four parameters $(w_{\mathrm{DE}}, q_{\mathrm{dec}}, j, s)$ 
from those for the $\Lambda$CDM model $(-1, -1, 1, 0)$ 
to be very small, so that this Pseudo-Rip model can be consistent with 
the $\Lambda$CDM model, similarly to that in the Little Rip model discussed in 
Sec.~IV C 3. 
%%%%%%%%
%%%%%%%%
In Table~\ref{tb:table3}, we display forms of $H$ and $f(T)$ with realizing 
(a) inflation in the early universe, (b) the $\Lambda$CDM model, 
(c) Little Rip cosmology and (d) Pseudo-Rip cosmology. 
%%%%%%%%

%%%%%%%%% Table III %%%%%%%%%%%
\begin{table*}[tbp]
\caption{Forms of $H$ and $f(T)$ with realizing 
%{\itshape (a)} 
(a) inflation in the early universe, (b) the $\Lambda$CDM model, 
(c) Little Rip cosmology and (d) Pseudo-Rip cosmology. 
}
\begin{center}
\begin{tabular}
{lll}
\hline
\hline
Cosmology
& $H$ 
& $f(T)$ 
\\[0mm]
\hline
(a) Power-law inflation
& $H = h_{\mathrm{inf}}/t$\,, 
%($h_{\mathrm{inf}} (> 1)$)
& $f(T) = A T^\alpha$\,, 
%where $\alpha <0$ or $\alpha = 1/2$
\\[0mm]
(In the limit of $t \to 0$)
& $h_{\mathrm{inf}} (> 1)$
& $\alpha <0$ or $\alpha = 1/2$
\\[0mm]
(b) $\Lambda$CDM model 
%(or exponential inflation)
& $H =\sqrt{\Lambda/3} = \mathrm{constant}$\,, \,\,\,\,\,
& $f(T) = T - 2\Lambda$\,, 
%where $\Lambda >0$
\\[0mm]
or exponential inflation
& $\Lambda >0$
& $\Lambda >0$
\\[0mm]
(c) Little Rip cosmology
& $H = H_{\mathrm{LR}} \exp \left( \xi t \right)$\,, 
%where $H_{\mathrm{lr}} (>0)$ and $\xi (>0)$
& $f(T) = A T^\alpha$\,, 
%where $\alpha <0$ or $\alpha = 1/2$
\\[0mm]
(In the limit of $t \to \infty$)
& $H_{\mathrm{LR}} > 0$ and $\xi >0$
& $\alpha <0$ or $\alpha = 1/2$
\\[0mm]
(d) Pseudo-Rip cosmology\,\,\,\,\,
& $H = H_{\mathrm{PR}} \tanh \left(t/t_0\right)$\,,
& $f(T) = A \sqrt{T}$ 
\\[0mm]
& $H_{\mathrm{PR}} > 0$
& 
\\[1mm]
\hline
\hline
\end{tabular}
\end{center}
\label{tb:table3}
\end{table*}
%%%%%%%%%%%%%%%%%%%%%%%%%%%%%%%%%%%%

In the expanding universe, the relative acceleration between two points 
separated by a distance $l$ is given by 
$l \ddot{a}/a$, where $a$ is the scale factor. 
Suppose that there exists a particle with mass $m$ at each of the points, 
an observer at one of the masses would measure an inertial force on the 
other mass. 
The inertial force $F_{\mathrm{inert}}$ on a mass $m$ is given 
by~\cite{Frampton:2011sp, Frampton:2011rh}
\begin{eqnarray}
F_{\mathrm{inert}} \Eqn{=} ml \frac{\ddot{a}}{a} 
= ml \left( \dot{H} + H^2 \right) 
\label{eq:FT4-6-IVC4-add-006} \\ 
\Eqn{=} -ml \frac{\kappa^2}{6} \left( \rho_{\mathrm{DE}} (a) 
+ 3 P_{\mathrm{DE}} (a) \right) 
= ml \frac{\kappa^2}{6} \left( 2\rho_{\mathrm{DE}} (a) 
+ \frac{d \rho_{\mathrm{DE}} (a)}{d a} a \right)\,,
\label{eq:FT4-6-IVC4-add-007} 
\end{eqnarray}
where in deriving the first equality in Eq.~(\ref{eq:FT4-6-IVC4-add-007}) 
we have used Eqs.~(\ref{eq:4.1}) and (\ref{eq:4.2}). 
We take the present value of $a$ as $a_0 \equiv a(t=t_0) = 1$. 
We also provide that the two particles are bound by a constant force 
$F_{\mathrm{b}}$. When $F_{\mathrm{inert}} (>0)$ is a positive force and 
the amplitude is larger than that of $F_{\mathrm{b}}$, the two particles 
become unbound and hence the bound structure is dissociated. 
In Pseudo-Rip cosmology, $F_{\mathrm{inert}}$ is asymptotically finite. 

For a Big Rip singularity realizing $H$ in Eq.~(\ref{eq:2.13}) 
with $q \geq 1$, by using Eq.~(\ref{eq:FT4-6-IVC4-add-006}) we find 
\begin{equation}
F_{\mathrm{inert}} 
= ml h_{\mathrm{s}} \left[ \frac{q}{\left( t_{\mathrm{s}} - t \right)^{q+1}} 
+ \frac{h_{\mathrm{s}}}{\left( t_{\mathrm{s}} - t \right)^{-2q}} \right] 
\longrightarrow \infty\,, 
\quad 
t \to t_{\mathrm{s}}\,. 
\label{eq:FT4-6-IVC4-add-008} 
\end{equation}
Here, 
the reason of the divergence is that $H$ and $\dot{H}$ diverge in the limit of 
$t \to t_{\mathrm{s}}$. 
Moreover. 
for a Little Rip model in Eq.~(\ref{eq:FT4-6-IVC3-01}), 
we see that 
\begin{equation}
F_{\mathrm{inert}} 
= ml H_{\mathrm{LR}} \left[ \xi + H_{\mathrm{LR}} \exp \left( \xi t \right) 
\right] \exp \left( \xi t \right)
\longrightarrow \infty\,, 
\quad 
t \to \infty\,. 
\label{eq:FT4-6-IVC4-add-009} 
\end{equation}
Similarly, the reason of the divergence is that 
$H$ and $\dot{H}$ diverge in the limit of $t \to \infty$. 
Thus, a phenomenon of ``Rip'' is produced by the cosmic accelerated expansion 
at a Big Rip singularity and in Little Rip cosmology. 

On the other hand, for a Pseudo-Rip model in 
Eq.~(\ref{eq:FT4-6-IVC4-add-005}), we obtain 
\begin{equation}
F_{\mathrm{inert}} 
= ml H_{\mathrm{PR}} \left[ 
\frac{1}{t_0 \cosh^2 \left( t/t_0 \right)} + H_{\mathrm{PR}} 
\tanh^2 \left( \frac{t}{t_0} \right)
\right] 
%\frac{1}{\cosh^2 \left( t/t_0 \right)} 
%\left[ \frac{1}{t_0} + H_{\mathrm{PR}} \sinh^2 \left( \frac{t}{t_0} \right) 
%\right] 
%\label{eq:FT4-6-IVC4-add-010} \\ 
\longrightarrow 
F_{\mathrm{inert}\,,\infty}^{\mathrm{PR}} 
%\equiv ml H_{\mathrm{PR}}^2 
< \infty\,, 
\quad 
t \to \infty\,, 
\label{eq:FT4-6-IVC4-add-010} 
\end{equation}
where 
\begin{equation}
F_{\mathrm{inert}\,,\infty}^{\mathrm{PR}} \equiv ml H_{\mathrm{PR}}^2\,. 
\label{eq:FT4-6-IVC4-add-011} 
\end{equation}
Therefore, $F_{\mathrm{inert}}$ is asymptotically finite. 
Here, the reason why $F_{\mathrm{inert}}$ becomes a finite value of 
$F_{\mathrm{inert}\,,\infty}^{\mathrm{PR}}$ in the limit of $t \to \infty$ 
is that $H \to H_{\mathrm{PR}}$ and $\dot{H} \to 0$. 

%%%%%
It is necessary for $F_{\mathrm{inert}\,,\infty}^{\mathrm{PR}}$ to be larger 
than the bound force $F_{\mathrm{b}}^{\mathrm{ES}} = G M_{\oplus} M_{\odot} / 
r_{\oplus-\odot}^2 = 4.37 \times 10^{16} \, 
\mathrm{GeV}^2$ of the ES system 
in order that the ES system will be disintegrated and hence 
the Pseudo-Rip scenario can be realized. 
Here, $M_{\oplus} = 3.357 \times 10^{51} \, 
\mathrm{GeV}$~\cite{Kolb and Turner} and $M_{\odot} = 1.116 \times 10^{57} \, 
\mathrm{GeV}$~\cite{Kolb and Turner} are masses of Earth and Sun, 
respectively, and $r_{\oplus-\odot} = 1 \mathrm{AU} = 7.5812 \times 10^{26} 
\, \mathrm{GeV}^{-1}$~\cite{Kolb and Turner} is the distance between 
Earth and Sun, i.e., the Astronomical unit. 
As an example, 
if we take $m$ as the Earth mass $m = M_{\oplus}$, 
$l$ is the distance between Earth and Sun 
$l = r_{\oplus-\odot}$, 
in order for $F_{\mathrm{inert}\,,\infty}^{\mathrm{PR}} > 
F_{\mathrm{b}}^{\mathrm{ES}}$, 
by using Eq.~(\ref{eq:FT4-6-IVC4-add-011}), we find $H_{\mathrm{PR}} > 
\sqrt{G M_{\odot} / r_{\oplus-\odot}^3} = 1.31 \times 10^{-31}\, 
\mathrm{GeV}$. If this condition is met, the disintegration of the ES system 
can occur much before arriving at de Sitter universe, 
so that the Pseudo-Rip scenario can be realized. 
In addition, if this constraint is satisfied, 
the current value of $w_{\mathrm{DE}}$ in this Pseudo-Rip model is also 
consistent with the observations because the constraint on $H_{\mathrm{PR}}$ 
from the current value of $w_{\mathrm{DE}}$ is much weaker as 
$H_{\mathrm{PR}} \geq 2.96 \times 10^{-42} \, \mathrm{GeV}$. 
%%%%%

%%%%%%%%%%%%%%%%%%%
%%%  Sec. IV D
%%%%%%%%%%%%%%%%%%%
\subsection{Time-dependent matter instability and star singularity in $f(T)$
gravity}

In modified gravity, the important property of the viability is the 
existence of the gravitational bound objects (stars, planets). 
It is checked via matter instability~\cite{Dolgov:2003px}. 
For instance, without such a property 
the relativistic star might not be formed 
because a corresponding singularity appears 
(for example, it is known that a singularity may appear for 
stars~\cite{Rel-stars} in $f(R)$ gravity). 
In a gravitating system with a time dependent mass density 
such as astronomical massive objects, 
the instability in $f(R)$ gravity has recently been 
studied~\cite{Arbuzova:2010iu}.  
Furthermore, 
the generation mechanism of the time-dependent matter instability 
in the star collapse has also been investigated. 
It has been demonstrated that the time-dependent matter instability develops 
and consequently the curvature singularity could appear~\cite{Bamba:2011sm}. 
In this subsection, by examining the process for the curvature singularity 
to be realized~\cite{Bamba:2011sm} in analogy with $f(R)$ gravity, 
we discuss whether the time-dependent matter instability in the star collapse 
occurs in $f(T)$ gravity. 

We examine a small region inside the star. 
We can regard this system as homogeneous and isotropic 
and hence the space-time is locally described by the flat FLRW metric 
(\ref{eq:2.8}). 
Here, the Hubble parameter $H$ is negative because 
we are exploring the star collapse, 
so that the space-time can be shrinking. 
Since the region is shrinking, 
the energy densities of the matters automatically increase. 
In case of cosmology, 
as shown in Sec.~III B, all the four types of the finite-time future 
singularities in $f(T)$ gravity can appear. 
If $H$ diverges in the limit of $t \to t_{\mathrm{st}}$, 
where $t_{\mathrm{st}}$ is the time when the time when the curvature 
singularity in the star appears, 
$T$ becomes infinity because $T = -6H^2$. 
{}From Table~\ref{tb:table1}, we see that 
for the Type III singularity, $T$ diverges, although the scale factor $a$ is 
finite. 
This phenomenon can be applied to the star collapse. 
The energy density and the pressure from the matter are 
finite and therefore these can be neglected near the singularity 
because $a$ asymptotically becomes a finite value. 
In this case, the Hubble parameter $H$ is expressed as 
\begin{equation}
H \sim - \frac{h_{\mathrm{st}}}{ \left( t_{\mathrm{st}} - t \right)^q}\,,
\label{eq:FT4-6-IVD-01} 
\end{equation}
where $h_{\mathrm{st}} (>0)$ is a positive constant. 
For the Type III singularity, we have $0<q<1$. 
In the limit of $t \to t_{\mathrm{st}}$, 
$T$ becomes infinite because $H$ diverges. 
This means that the naked curvature singularity appears 
in the finite future. 
We note that in the above investigations, we have supposed that 
the region is almost homogeneous and isotropic. 
When these settings are adequate 
also near the singularity, the singularity simultaneously occurs 
in all of the region. 
The naked singularities occur densely in the region, 
even though the homogeneity and isotropy are broken. 
Moreover, the density of matter grows as 
in the region farther from the surface of the star, namely, 
nearer to the center of it. 
Accordingly, first the naked curvature singularity 
can appear near the center of the star. 
When the singularity produces the attractive force, 
the shrinking of the star proceeds more and more, 
whereas when the repulsive force is generated, 
eventually the explosion may happen. 
However, 
when the explosion happens, the sign of $H$ has to change 
from negative to positive. The realization of this phenomenon 
seems to be difficult. 

On the other hand, in the cosmological context, 
when the Hubble parameter $H$ is given by 
Eq.~(\ref{eq:2.13}) with $0<q<1$, the Type III singularity can appear. 
By using the initial condition for $H$ and $\dot{H}$, we can determine 
the values of $t_{\mathrm{s}}$ (and $h_{\mathrm{s}}$) in Eq.~(\ref{eq:2.13})
and $t_{\mathrm{st}}$ (and $h_{\mathrm{st}}$) in 
Eq.~(\ref{eq:FT4-6-IVD-01}). 
The absolute values of $H$ and $\dot{H}$ inside the star 
would be larger than those in the expanding universe 
because we are investigating the collapsing star. 
Hence, we find $t_{\mathrm{s}} > t_{\mathrm{st}}$. 
In other words, 
the curvature singularity in the star appears before the cosmological 
singularity. 
As a consequence, 
the time-dependent matter instability in the star collapse 
can occur in $f(T)$ gravity, similarly to that in $f(R)$ gravity.

%%%%%%%%%%%%%%%%%%%
%%%  Sec. V
%%%%%%%%%%%%%%%%%%%
\section{Thermodynamics around the finite-time future singularities}

In this section, 
we discuss thermodynamics in $f(T)$ gravity in order to examine whether 
$f(T)$ gravity is a viable gravitational theory. 
In particular, by following the procedure 
in Refs.~\cite{Bamba:2011pz, Bamba:2010kf}, 
we examine whether the second law of thermodynamics can be 
verified around the finite-time future singularities . 
%%%%%%%%
The fundamental connection between gravitation and thermodynamics 
was implied by Black hole thermodynamics~\cite{BCH-B-H} 
(for recent reviews, see, e.g.,~\cite{Rev-T-G-Pad}). 
In general relativity, 
by using the proportionality of the entropy to the horizon 
area, the Einstein equation was derived from 
the Clausius relation in thermodynamics~\cite{Jacobson:1995ab}. 
This consequence has been app lied to 
more general extended gravitational 
theories~\cite{Modified-gravity-EOS, Brustein-Hadad-Medved}. 
%%%%%%%%

%%%%%%%%%%%%%%%%%%%%%%%%%%%
%%%  Sec. V A
%%%%%%%%%%%%%%%%%%%%%%%%%%%
\subsection{First law of thermodynamics}

In Refs.~\cite{Bamba:2010kf, Bamba:2011pz, BGT}, it has been shown that 
if the standard continuity equation (\ref{eq:4.5}) in terms of 
dark component is satisfied, an equilibrium description of thermodynamics 
can be obtained. 
In the flat FLRW space-time, the radius $\tilde{r}_A$ of 
the apparent horizon is written by 
%
%\begin{equation}
$
\tilde{r}_A= 1/H
$.
%\label{eq:VA-3.6}
%\end{equation}
%
The dynamical apparent horizon is determined by the relation 
$h^{\alpha \beta} \partial_{\alpha} \tilde{r} \partial_{\beta} \tilde{r}=0$. 
The time derivative of $\tilde{r}_A= 1/H$ leads to 
%
%\begin{equation}
$
-d\tilde{r}_A/\tilde{r}_A^3 
=\dot{H}H dt
$. 
%\label{eq:VA-3.7}
%\end{equation}
%
By plugging Eq.~(\ref{eq:4.1}) with this equation, we find the relation 
%
%\begin{equation} 
$
\left[1/\left(4\pi G\right)\right] d\tilde{r}_A=\tilde{r}_A^3 H
\left( \rho_{\mathrm{t}}+P_{\mathrm{t}} \right) dt
$. 
%\label{Fdr2}
%\label{eq:VA-4.6}
%\end{equation} 
%
Here, 
$\rho_{\mathrm{t}} \equiv \rho_{\mathrm{DE}}+
\rho_{\mathrm{M}}$ and $P_{\mathrm{t}} \equiv P_{\mathrm{DE}}+P_{\mathrm{M}}$ 
are the total energy density and pressure of the universe, respectively. 
In general relativity, the Bekenstein-Hawking horizon entropy is 
expressed as 
% 
%\begin{equation} 
$
S=\mathcal{A}/\left(4G\right)
$ 
%\label{Sdef2}
%\label{eq:VA-4.7}
%\end{equation} 
%
with $\mathcal{A}=4\pi \tilde{r}_A^2$ being the area of the apparent 
horizon~\cite{BCH-B-H}. 
By using the horizon entropy 
%in Eq.~(\ref{eq:VA-4.7}) 
and the above relation,    
%using Eq.~(\ref{eq:VA-4.6}), 
we acquire  
\begin{equation} 
\frac{1}{2\pi \tilde{r}_A} dS=4\pi \tilde{r}_A^3 H
\left( \rho_{\mathrm{t}}+P_{\mathrm{t}} \right) dt \,.
%\label{dSre2}
\label{eq:VA-4.8}
\end{equation} 
%
%The associated temperature of the apparent horizon has 
The Hawking temperature 
%$T_{\mathrm{H}}$ 
%
%\begin{equation}
$
T_{\mathrm{H}} = |\kappa_{\mathrm{sg}}|/\left(2\pi\right)
$ 
corresponds to the associated temperature of the apparent horizon. 
%\label{eq:VA-3.11} 
%\\
%\end{equation}
%
Here, the surface gravity $\kappa_{\mathrm{sg}}$ is 
represented by~\cite{Cai:2005ra} 
\begin{eqnarray}
\kappa_{\mathrm{sg}} 
\Eqn{=} 
\frac{1}{2\sqrt{-h}} \partial_\alpha 
\left( \sqrt{-h}h^{\alpha\beta} \partial_\beta \tilde{r} \right) 
\label{eq:VA-3.12} \\
\Eqn{=} -\frac{1}{\tilde{r}_A}
\left( 1-\frac{\dot{\tilde{r}}_A}{2H\tilde{r}_A} \right)
=-\frac{\tilde{r}_A}{2} \left( 2H^2+\dot{H} \right) 
= 
-\frac{2\pi G}{3F} \tilde{r}_A 
\left( \rho_{\mathrm{t}}-3P_{\mathrm{t}} \right) 
\label{eq:VA-3.13} \\
\Eqn{=} 
-\frac{2\pi G}{3F} \tilde{r}_A 
\left(1 -3w_{\mathrm{t}} \right) \rho_{\mathrm{t}}\,, 
\label{eq:FT4-13-VA-add-1}
\end{eqnarray}
where $h$ is the determinant of the metric $h_{\alpha\beta}$ 
and $w_{\mathrm{t}} \equiv P_{\mathrm{t}}/\rho_{\mathrm{t}}$ is 
the EoS for the total of energy and matter in the universe. 
%%%%%
It follows from Eq.~(\ref{eq:FT4-13-VA-add-1}) that for 
$w_{\mathrm{t}} \le 1/3$, we have 
$\kappa_{\mathrm{sg}} \le 0$. 
%%%%%
Thus, by substituting Eq.~(\ref{eq:VA-3.13}) into 
$T_{\mathrm{H}} = |\kappa_{\mathrm{sg}}|/\left(2\pi\right)$, 
we obtain 
\begin{equation}
T_{\mathrm{H}}=\frac{1}{2\pi \tilde{r}_A}
\left( 1-\frac{\dot{\tilde{r}}_A}{2H\tilde{r}_A} \right)\,.
%\label{tempe}
\label{eq:VA-3.14}
\end{equation}
By combining 
Eq.~(\ref{eq:VA-4.8}) with Eq.~(\ref{eq:VA-3.14}), we acquire 
\begin{equation} 
T_{\mathrm{H}} dS = 4\pi \tilde{r}_A^3 H \left(\rho_{\mathrm{t}}+P_{\mathrm{t}} \right) dt 
-2\pi  \tilde{r}_A^2 \left(\rho_{\mathrm{t}}+P_{\mathrm{t}} \right) 
d\tilde{r}_A\,.
%\label{TdS2}
\label{eq:VA-4.9}
\end{equation} 
The Misner-Sharp energy~\cite{Misner-Sharp-energy} is defined as 
%
%\begin{equation} 
$
E=\tilde{r}_A/\left(2G\right) = 
V\rho_{\mathrm{t}}
$ 
%\label{Misner2}
%\label{eq:VA-4.10}
%\end{equation} 
%
with $V=4\pi \tilde{r}_A^3/3$ being the volume inside 
the apparent horizon. The last equality 
implies that $E$ is equivalent to the total intrinsic energy. 
By using this equation, we have 
\begin{equation} 
dE=-4\pi \tilde{r}_A^3 H \left(\rho_{\mathrm{t}}+P_{\mathrm{t}} \right) dt 
+4\pi \tilde{r}_A^2 \rho_{\mathrm{t}} d\tilde{r}_A\,.
%\label{dE2}
\label{eq:VA-4.11}
\end{equation} 
We also define the work density~\cite{W-D}: 
%given by 
%
%\begin{eqnarray} 
$
W \equiv
-\left (1/2\right) \left( T^{(\mathrm{M})\alpha\beta}
h_{\alpha\beta} + T^{(\mathrm{DE})\alpha\beta} h_{\alpha\beta} 
\right) 
%\label{eq:VA-3.20} \\ 
%\Eqn{=} 
=
%\frac{1}{2} 
\left (1/2\right) 
\left( \rho_{\mathrm{t}}-P_{\mathrm{t}} \right) 
%\label{work2}
%\label{eq:VA-4.13}
%\end{eqnarray} 
%
$. 
Here, $T^{(\mathrm{M})\alpha\beta}$ and $T^{(\mathrm{DE})\alpha\beta}$ are 
the energy-momentum tensor of matter and that of dark components, 
respectively. 
By using Eq.~(\ref{eq:VA-4.11}) and the work density $W$, 
the first law of equilibrium thermodynamics can be described. 
\begin{equation} 
T_{\mathrm{H}} dS=-dE+W dV
%\label{first}
\label{eq:VA-4.12}
\end{equation} 
As a result, an equilibrium description of thermodynamics can be realized. 
We also remark that 
from Eqs.~(\ref{eq:4.1}), (\ref{eq:4.2}) and (\ref{eq:VA-4.8}), 
we have 
%
%\begin{equation}
$
\dot{S} = 
8\pi^2 H \tilde{r}_A^4 \left(\rho_{\mathrm{t}}+P_{\mathrm{t}}\right) 
= \left( 6\pi/G \right) \left( \dot{T}/T^2 \right) 
= -\left( 2\pi/G \right) \left[ \dot{H}/\left( 3H^3 \right) \right] 
> 0
$. 
%\label{dSre3}
%\label{eq:VA-4.14}
%\end{equation}
%
This means that in the expanding universe, the horizon entropy $S$ always 
increases as long as the null energy condition 
$\rho_{\mathrm{t}}+P_{\mathrm{t}} 
\ge 0$, i.e., $\dot{H} \leq 0$, is met. 
%%%%%

%%%%%%%%%%%%%%%%%%%%%%%%%%%
%%%  Sec. IV B
%%%%%%%%%%%%%%%%%%%%%%%%%%%
\subsection{Second law of thermodynamics}

Next, we investigate the second law of thermodynamics in the equilibrium 
description. The Gibbs equation in terms of all matter and energy 
fluid is represented by 
%
%\begin{equation}
$
T_{\mathrm{H}} dS_{\mathrm{t}} = d\left( \rho_{\mathrm{t}} V \right) +P_{\mathrm{t}} dV 
= V d\rho_{\mathrm{t}} + \left( \rho_{\mathrm{t}} +P_{\mathrm{t}} 
\right) dV
$. 
%\label{eq:VB-4.A001}
%\end{equation}
%
The second law of thermodynamics can be expressed as 
%
%\begin{equation}
%
$
dS_{\mathrm{sum}}/dt
\equiv 
dS/dt + dS_{\mathrm{t}}/dt 
\geq 0
$
%
%\label{eq:VB-4.A002}
%\end{equation}
%
with $S_{\mathrm{sum}} \equiv S + S_{\mathrm{t}}$, 
where $S_{\mathrm{t}}$ is the entropy of total energy inside the 
horizon. 
By using $V=4\pi \tilde{r}_A^3/3$, Eqs.~(\ref{eq:4.2}), 
(\ref{eq:VA-3.14}) and the relation 
$
\dot{S} = 
8\pi^2 H \tilde{r}_A^4 \left(\rho_{\mathrm{t}}+P_{\mathrm{t}}\right) 
= \left( 6\pi/G \right) \left( \dot{T}/T^2 \right)  
$, 
we acquire 
\begin{equation}
\frac{dS_{\mathrm{sum}}}{dt} 
= 
-\frac{6\pi}{G}\left( \frac{\dot{T}}{T} \right)^2 \frac{1}{4HT + \dot{T}}\,.
\label{eq:VB-4.A003}
\end{equation}
As a result, from the second law of thermodynamics with 
Eq.~(\ref{eq:VB-4.A003}) we find the condition~\cite{Bamba:2011pz} 
%
%\begin{equation}
$
Y \equiv -\left( 4HT + \dot{T} \right) 
= 12H \left( 2H^2 + \dot{H} \right) 
\geq 0
$. 
%\label{eq:VB-4.A004}
%\end{equation}
%
%%%%%
{}From the behaviors of $H$ and $\dot{H}$ in the limit of 
$t \to t_{\mathrm{s}}$ described in Table~\ref{tb:table1}, it can be seen that 
in the expanding universe, where $H>0$, 
for all the four types of the finite-time future singularities, 
the relation $2H^2 + \dot{H} \geq 0$ can always be realized. 
It is also interesting to note that this relation is satisfied 
even in the phantom phase ($\dot{H} >0$). 
Thus, the second law of thermodynamics around the finite-time future 
singularities is always satisfied. 
%%%%% 
%%%%%%%%
However, it should be cautioned that 
at the exact singularity $t=t_{\mathrm{s}}$ such as a Big Rip singularity, 
in which the scale factor $a (t)$ diverges as $a (t=t_{\mathrm{s}}) = \infty$, 
a naive classical picture of thermodynamics might break down. 
%%%%%%%%
We also mention that this result can be verified only provided that 
the temperature of the universe inside the apparent 
horizon is equal to that of the horizon~\cite{GWW-JSS}.

%%%%%%%%%%%%%%%%%%%
%%%  Sec. VI
%%%%%%%%%%%%%%%%%%%
\section{Conclusions}

In the present paper, 
we have reconstructed $f(T)$ gravity 
in which finite-time future singularities appear. 
Furthermore, it has been demonstrated that 
a $T^\beta$ ($\beta > 1$) correction term, e.g., $\beta = 2$, to 
a model of $f(T)$ gravity 
where the finite-time future singularities occur, 
can remove the finite-time future singularities, 
similarly to that in $f(R)$ gravity. 
We have also explored non-singular $f(T)$ models in which 
inflation in the early universe, 
the $\Lambda$CDM model, Little Rip cosmology and Pseudo-Rip cosmology 
are realized. 
It has been shown that the dissolution of bound structures for 
Little Rip and Pseudo-Rip cosmologies happens in the same manner as in 
gravity with corresponding dark energy fluid. 
Moreover, we have considered that the time-dependent matter instability 
in the star collapse can occur in $f(T)$ gravity, similarly to that 
in $f(R)$ gravity. 
In addition, we have investigated thermodynamics in $f(T)$ gravity and 
shown that the second law of thermodynamics can be satisfied around the 
finite-time future singularities, provided that the temperature of the universe inside the horizon is equal to that of the apparent horizon. 

%%%%%%%%
It would be interesting to develop more complicated versions of $f(T)$ gravity 
and study their cosmological applications. 
For instance, one can consider non-minimal coupling of $f_1 (T)$ with 
electrodynamics, where $f_1 (T)$ is a function of $T$, 
in analogy with that in non-minimal $f_1 (R)$ gravity~\cite{Bamba:2011nm}. 
This may lead to the emergence of a domain-wall solution and 
variation of fine structure constant in non-minimal $f_1 (T)$ gravity. 
{}From another side, it seems to be very interesting to generalize an $f(T)$ 
theory in a consistent way so that one can include the presence of 
curvature in the Lagrangian. 
%%%%%%%%

It should be emphasized that the illustration of 
the existence of the finite-time future singularities 
in $f(T)$ gravity and the possibility of those removing due to 
an additional power-low term are nontrivial and significant. 
These consequences are also found in other alternative gravitational theories 
such as $f(R)$ gravity. 
It is considered that the removal possibility of the finite-time future 
singularities can be one of the tests of a successful alternative 
gravitational theory to general relativity. 
Therefore, it is strongly expected that 
through these analyses of meaningful theoretical properties of modified 
gravitational theories, we can obtain a successful alternative gravitational 
theory to general relativity which explains the cosmic accelerated expansion 
of the universe in a geometrical way.

%%%%%%%%%%%%%%%%%%%%%%%%
%%%  Acknowledgments
%%%%%%%%%%%%%%%%%%%%%%%%
\section*{Acknowledgments}

%%%
S.D.O. would like to 
appreciate the support and very kind hospitality 
at Eurasian National University 
and also acknowledge the Japan Society for the Promotion of Science (JSPS) 
Short Term Visitor Program S11135 
and the very warm hospitality at 
Nagoya University where the work was developed. 
%%%
The work is supported in part
by Global COE Program
of Nagoya University (G07) provided by the Ministry of Education, Culture,
Sports, Science \& Technology
(S.N.); 
the JSPS Grant-in-Aid for Scientific Research (S) \# 22224003 and (C) 
\# 23540296 (S.N.); 
and
MEC (Spain) project FIS2010-15640 and AGAUR (Catalonia) 2009SGR-994
(S.D.O.).

%%%%%%%%%%%%%%%%%%%
%%%  Appendix
%%%%%%%%%%%%%%%%%%%
\appendix
%%%%%%%%%%%%%%%%%%%
%%%  A.
%%%%%%%%%%%%%%%%%%%
\section{Reconstruction method}

In this appendix, we explain the procedure of a reconstruction method of 
$f(T)$ gravity 
by applying that of $f(R)$ gravity~\cite{Reconstruction-Method, 
%Nojiri:2006gh, Nojiri:2006be, Nojiri:2008fk, 
Nojiri:2009kx, Nojiri:2011kd} to $f(T)$ gravity. 
The action of $f(T)$ gravity with matter is given by Eq.~(\ref{eq:2.6}). 
We can rewrite the action in Eq.~(\ref{eq:2.6}) with proper functions 
$P(\phi)$ and $Q(\phi)$ of a scalar field $\phi$ to 
\begin{equation}
S=\int d^4 x \sqrt{-g} 
%\left\{ 
\frac{1}{2\kappa^2} \left( P(\phi) T + Q(\phi) \right) 
+ \int d^4 x 
{\mathcal{L}}_{\mathrm{M}} \left( g_{\mu\nu}, {\Psi}_{\mathrm{M}} \right)
%+ {\mathcal{L}}_{\mathrm{matter}} 
%\right\}
\,.
\label{eq:VIII-B-2}
\end{equation}
Since the scalar field $\phi$ does not have kinetic term, we may regard 
$\phi$ as an auxiliary scalar field. 
{}From Eq.~(\ref{eq:VIII-B-2}), we derive the equation of 
motion of $\phi$ as 
%
%\begin{equation}
$
0=\left( d P(\phi)/d \phi \right ) T + d Q(\phi)/d \phi
$. 
%\label{eq:VIII-B-3}
%\end{equation}
%
The substitution of $\phi=\phi(T)$ into the action in Eq.~(\ref{eq:VIII-B-2}) 
leads to the expression of $f(T)$ as 
%
%\begin{equation}
$
f(T) = P(\phi(T)) T + Q(\phi(T))
$.
%\label{eq:VIII-B-4}
%\end{equation}
%
It follows from Eq.~(\ref{eq:2.7}) that 
the gravitational field equation is described by 
\begin{eqnarray}
&&
\frac{1}{e} \partial_\mu \left( eS_A^{\verb| |\mu\nu} \right) P(\phi) 
-e_A^\lambda T^\rho_{\verb| |\mu \lambda} S_\rho^{\verb| |\nu\mu} 
P(\phi) +S_A^{\verb| |\mu\nu} \partial_\mu \left(T\right) 
\frac{d P(\phi)}{d \phi} \frac{d \phi}{d T} 
\nonumber \\
&&
\hspace{50mm}
{}+\frac{1}{4} e_A^\nu \left( P(\phi) T + Q(\phi) \right) 
= \frac{{\kappa}^2}{2} e_A^\rho 
{T^{(\mathrm{M})}}_\rho^{\verb| |\nu}\,.
\label{eq:VIII-B-5}
\end{eqnarray}
%

%%%%%%%%
In the flat FLRW background space-time with the metric in Eq.~(\ref{eq:2.8}), 
the components of $(\mu,\nu)=(0,0)$ and $(\mu,\nu)=(i,j)$ $(i,j=1,\cdots,3)$ 
in Eq.~(\ref{eq:VIII-B-5}) yield two independent differential equations in 
terms of $P(\phi(t))$ and $Q(\phi(t))$. In principle, by eliminating $Q(\phi)$ 
from these two equations, we obtain an equation which $P(\phi(t))$ obeys. 
We may take the scalar field $\phi$ as $\phi = t$ by redefining it properly. 
We express $a(t)$ as
%
%\begin{equation}
$
a(t) = \bar{a} \exp \left( \tilde{g}(t) \right)
$, 
%\label{eq:VIII-B-10}
%\end{equation}
%
where $\bar{a}$ is 
%in terms of 
a constant and $\tilde{g}(t)$ is a proper function of $t$. 
By using $H= d \tilde{g}(\phi)/\left(d \phi \right)$, we represent 
the equation in terms of $P(\phi(t))$ derived above 
to be a form described by $P(\phi(t))$, the derivatives of $P(\phi(t))$ 
in terms of $\phi$, $d \tilde{g}(\phi)/\left(d \phi \right)$, and the 
derivatives of $d \tilde{g}(\phi)/\left(d \phi \right)$ 
in terms of $\phi$. 
Furthermore, from another equation we can also express 
$Q(\phi(t))$ with $P(\phi(t))$, the derivatives of $P(\phi(t))$ 
in terms of $\phi$, $d \tilde{g}(\phi)/\left(d \phi \right)$, and the 
derivatives of $d \tilde{g}(\phi)/\left(d \phi \right)$ 
in terms of $\phi$. 
We derive the solutions of $P(\phi(t))$ and $Q(\phi(t))$ and substitute 
these solutions to $f(T) = P(\phi(T)) T + Q(\phi(T))$, 
we acquire the expression of $f(T)$ as a function of only $T$. 
%%%%%%%%

%%%%%
Finally, we remark the following point. 
By redefining the auxiliary scalar field $\phi$
as $\phi \equiv \Phi(\varphi)$ with a proper function $\Phi$ and defining 
$\tilde P(\varphi)\equiv P(\Phi(\varphi))$ and
$\tilde Q(\varphi)\equiv Q(\Phi(\varphi))$, the new action 
%
%\begin{eqnarray}
$
S = \int d^4 x \sqrt{-g}
\tilde f(T)/\left( 2\kappa^2 \right)
+ \int d^4 x 
{\mathcal{L}}_{\mathrm{M}} \left( g_{\mu\nu}, {\Psi}_{\mathrm{M}} \right)
$, 
where 
%\label{eq:VIII-B-A1} \\
$
\tilde f(T) \equiv \tilde P(\varphi) T + \tilde Q(\varphi)
$, 
%\label{eq:VIII-B-A2}
%\end{eqnarray}
%
is equivalent to the action in Eq.~(\ref{eq:VIII-B-2}). 
This is because $\tilde f(T) = f(T)$. 
Since $\varphi$ is the inverse function of $\Phi$, 
by using $\phi = \phi(T)$, 
$\varphi$ can be solved with respect to $T$ as 
$\varphi=\varphi(T) = \Phi^{-1}(\phi(T))$. 
Accordingly, there exist the choices in $\phi$ as a gauge symmetry 
and therefore $\phi$ can be identified with time $t$ as $\phi=t$. 
We can interpret this fact as a gauge condition which is equivalent to 
the reparameterization of 
$\phi=\phi(\varphi)$~\cite{BGNO-BG}. 
As a result, 
we can find the form of $f(T)$ by obtaining the relation $t = t(T)$. 
In addition, 
regarding the relation between $H$ and $T$ in $f(T)$ gravity and 
that between $H$ and $R$ in $f(R)$ gravity 
in the flat FLRW background space-time, 
we note that for $f(T)$ gravity and $f(R)$ gravity, we have $T=-6H^2$ and 
$R = 6\left(2H^2 + \dot{H} \right)$, respectively, and therefore 
in comparison with $f(R)$ gravity, in $f(T)$ gravity the torsion scalar $T$ 
depends on only $H$, although in $f(R)$ gravity the scalar curvature $R$ 
depends on both $H$ and $\dot{H}$. 
%%%%%

%%%%%%%%%%%%%%%%%%%%%%%%%%%%%%%%%
%% thebibliography environment
%%%%%%%%%%%%%%%%%%%%%%%%%%%%%%%%%

\end{document}